\documentclass[useAMS,usenatbib]{mn2e}

\usepackage{epsfig}

\usepackage{amsmath,amssymb,bm} 

\title[Tidal Warping of Be Star Decretion Discs]{Tidal Warping and Precession of Be
  Star Decretion Discs}
  
\author[R. G. Martin et al.]{Rebecca G.  Martin$^1$\thanks{E-mail:
    rmartin@stsci.edu}, J. E. Pringle$^2$, Christopher A. Tout$^2$ and
  Stephen H. Lubow$^1$ \\$^1$Space Telescope Science Institute, 3700 San
  Martin Drive, Baltimore, MD 21218, USA \\$^2$University of
  Cambridge, Institute of Astronomy, The Observatories, Madingley
  Road, Cambridge, CB3 0HA, UK\\}
\begin{document}

\date{}

\pagerange{\pageref{firstpage}--\pageref{lastpage}} 
\pubyear{2010}
\maketitle

\label{firstpage}

\begin{abstract}
Rapidly rotating Be stars are observed as shell stars when the
decretion disc is viewed edge on. Transitions between the two implies
that the discs may be warped and precessing. Type~II X-ray outbursts
are thought to occur when the warped disc interacts with the fast
stellar wind.  We suggest that tides from a misaligned companion
neutron star can cause the observed effects. We make numerical models
of a Be star decretion disc in which the spin of the Be star is
misaligned with the orbital axis of a neutron star companion.  Tidal
torques from the neutron star truncate the disc at a radius small
enough that the neutron star orbit does not intersect the disc unless
the eccentricity or misalignment is very large. A magnetic torque from
the Be star that is largest at the equator, where the rotation is
fastest, is approximated by an inner boundary condition.  There are
large oscillations in the mass and inclination of the disc as it moves
towards a steady state. These large variations may explain the
observed changes from Be star to Be shell star and vice-versa and also
the Type~II X-ray outbursts. We find the tidal timescale on which the
disc warps, precesses and reaches a steady state to be around a year
up to a few hundred years. If present, the oscillations in mass and
disc inclination occur on a fraction of this timescale depending on
the orbital parameters of the binary. The timescales associated with
the tidal torque for observed Be star binaries suggest that these
effects are important in all but the longest period binaries.
\end{abstract}


\section{Introduction}

Be stars are rapidly rotating close to their break up velocity
\citep{Porter96}.  They are early type main-sequence stars which have
shown H$\alpha$ in emission at least once. They are variable in
brightness and spectra which show broad HeI absorption and emission at
either visual or UV wavelengths.  It is found that the emission, and
so presumably their discs, are only temporary and so Be stars become B
stars and vice versa. Be star discs form and disappear on a timescale
of a few months to a few years \citep{telting, bjork00, bjork02}. More
recently, \cite{raj10} observed the optical emission from Be X-ray
binaries in the SMC and found superorbital quasi-periodic variations
on timescales of about $200-3000\,\rm d$ that are thought to be related
to the formation and depletion of the circumstellar discs.

We choose to model the Be star disc as a viscous decretion disc.
Several other models have previously been proposed to describe the
physical structure of the disc such as the wind-compressed disc model
and the magnetically wind-compressed disc model \citep[see][for a
  review]{porter03}. In the wind-compressed disc model the rotation of
the star produces a wind towards the equator. The material from both
sides collides at the equator and the shock produces a dense region
confined to the equatorial plane by the opposing winds from each side
\citep{bjorkman93}. However, non-radial effects because of the
rotation of the star ruled out this picture from dynamical
arguments. It cannot reproduce the observed IR excess or kinematic
structure \citep{porter97}. An attempt to revive this model was made
with the addition of magnetic fields \citep{cassinelli00}. However,
numerical simulations of this model still do not explain the
observations when stellar rotation is included \citep{owoki03}. The
most likely alternative to the wind-compressed disc model is the
viscous disc model that we have used in this work.

The disc is modelled as a viscous decretion discs with the mass
expelled from the neighbourhood of the Be~star itself
\citep{lee91,cass02}.  \cite{han} finds that the discs are
rotationally supported and \cite{quirr} deduced that the circumstellar
envelopes around Be stars are thin discs from interferometry of
H$\alpha$ emission.  \cite{wood97} showed that the discs have an
opening angle of about $2.5^\circ$ with polarimetry and this is
consistent with a Keplerian disc. This is equivalent to an aspect
ratio $H/R\approx 0.04$.

Decretion discs have a different structure to accretion discs because
the inner boundary condition is different \citep{pringle91}. In an
accretion disc the inner boundary has zero torque and accretion is
allowed on to the central object. In a decretion disc, the inner
boundary prevents flow there on to the central object. The disc grows
when the star exerts a positive torque on it or decays when there is
no torque.  If mass is injected on to an accretion disc, then outside
of the injection radius, the disc is a decretion disc. Continual
addition of angular momentum allows the viscosity to transport angular
momentum outwards though the disc and the disc mass moves outwards.
There are mathematical disc solutions for the surface density of
decretion discs which either have infinite mass or divergent radial
velocity \citep{Okazaki07}.  There is a third steady state decretion
disc solution where there is no flow either in or out. Flow may be
prevented in the outer regions of a disc by a torque, such a tides
from a companion star and in the inner regions by a magnetic torque
from the central star.

If the disc is viewed edge on the Be~star is seen as a shell star.
Two Be stars, $\gamma$ Cas and 59~Cyg have shown two successive shell
events. which have been explained by a disc that is tilted with
respect to the equatorial plane of the star \citep{Hummel98}. The
variation in emission line widths and profile shapes are then due to
the precession of the disc.  The idea that a disc might change its
inclination to the line of sight is further investigated through
observations of 28 Tau (Pleione) by \cite{hirata07}. This star also
changes between B star, Be star and shell star on a $35\,\rm yr$
timescale and the intrinsic polarization angle changes in phase with
these variations.  The cause of the precession is also not clear but
\cite{Hummel98} suggested that it might be induced by tides from a
binary companion as most Be stars have a neutron star
companion.

Be/X-ray binaries show three types of X-ray activity
\citep{stella86,neg98}. Type~I outbursts are the periodic X-ray
outbursts that coincide with periastron passage with a luminosity
$L_X=10^{36}-10^{37}\,\rm erg\,s^{-1}$. Type~II outbursts last weeks
to months and show no orbital modulation but have a larger luminosity
of $L_X>10^{37}\,\rm erg\,s^{-1}$. It has been suggested that the
Type~II outbursts are linked to the interaction of the fast stellar
wind with a warped disc \citep{neg01,okazaki01}.  The third type of
X-ray activity is persistent low luminosity X-ray emission with
$L_X<10^{34}\,\rm erg\,s^{-1}$.  In this work we consider the warping
and precession of Be star discs with a neutron star companion on an
orbit that is misaligned with the spin of the Be star and attempt to
explain this activity.

In Section~\ref{structure} we consider general properties of accretion
and decretion disc structure.  In Section~\ref{tidal} we consider
properties of a decretion disc model warped by a tidal torque and in
Section~\ref{numerical} we make numerical models. In
Section~\ref{applications} we apply the models to observed Be and B
star binaries and consider how they are affected by a tidal torque.

\section{Disc Structure}
\label{structure}

We model the structure and evolution of warped accretion discs
according to the simplified formalism described by \cite{pringle92}.
We consider the disc to be made up of annuli of width $dR$ and mass
$2\pi \Sigma R dR$ at radius $R$ from the central Be star of mass
$M_1$ with surface density $\Sigma(R,t)$ at time $t$. The material is
in Keplerian orbits with angular velocity
$\Omega=\sqrt{GM_1/R^3}$. The angular momentum at radius $R$ is
\begin{equation}
\bm{L}=(GM_1R)^{1/2}\Sigma \bm{l}=L\bm{l},
\end{equation}
 where
$\bm{l}=(l_x,l_y,l_z)$ is a unit vector describing the direction of
the angular momentum of a disc annulus with $|\bm{l}|=1$.

We use equation (2.8) of \cite{pringle92} and add the binary tidal
torque, $\bm{T}_{\rm tid}$, and the angular momentum source term,
$\bm{\dot L}_{\rm acc}$, to give
\begin{align}
  \frac{\partial \bm{L}}{\partial
    t}=\,&\frac{1}{R}\frac{\partial}{\partial R}\left[ \left(
      \frac{3R}{L} \frac{\partial}{\partial R}(\nu_1 L)
      -\frac{3}{2}\nu_1\right)\bm{L}+\frac{1}{2}\nu_2RL\frac{\partial
      \bm{l}}{\partial R}\right] \cr & + \frac{1}{R}
\frac{\partial}{\partial R} \left( \nu_2 R^2 \left| \frac{\partial
      \bm{l}}{\partial R} \right|^2 \bm{L}\right)  
+\bm{T}_{\rm tid}+\bm{\dot L}_{\rm acc}.
\label{maineq}
\end{align}
The source of angular momentum by mass accretion on to the disc from the Be
star is
\begin{equation}
 \bm{\dot L}_{\rm acc} = \frac{\dot M}{2\pi R_{\rm add}} R_{\rm add}^2\Omega_{\rm add}\,  \delta (R-R_{\rm add}) \,  \bm{j}_{\rm Be},
\label{add}
\end{equation}
where
\begin{equation}
\bm{j}_{\rm Be}=(\sin\theta,0,\cos \theta)
\label{jadd}
\end{equation}
is the direction of the spin of the Be star. We assume the material
added has angular momentum aligned with the Be star spin and it is
injected at a radius $R_{\rm add}$ so that $\Omega_{\rm
  add}=\Omega(R_{\rm add})$. We work in the frame where the binary
orbital axis is $(0,0,1)$. We discuss the tidal torque in
Section~\ref{tidal}.

There are two viscosities. First $\nu_1$ corresponds to the azimuthal
shear, the viscosity normally associated with accretion discs, and
secondly $\nu_2$ corresponds to the vertical shear in the disc which
smoothes out the twist. The second viscosity acts when the disc is
non-planar.  With the $\alpha$-parametrisation the viscosities are
given by 
\begin{equation}
\nu_i=\alpha_i c_{\rm s} H
\label{viscs}
\end{equation}
 \citep{shakura73}, where $i=1,2$, $c_{\rm s}=H\Omega$ is the sound
 speed and $H$ is the scale height of the disc. The dimensionless
 parameters $\alpha_1 \le 1$ and $\alpha_2$ must be determined
 experimentally. \cite{lee91} suggest that with $\alpha=0.1$ it will
 take too long for the disc mass to build up so we choose
 $\alpha_1=0.3$ \citep[e.g.][]{jones08}. 
\cite{ogilvie99} found the viscosities to be related by
\begin{equation}
\frac{\nu_2}{\nu_1}=\frac{\alpha_2}{\alpha_1}=\frac{1}{2\alpha_1^2}\frac{4(1+7\alpha_1^2)}{4+\alpha_1^2}
\end{equation}
and more recently \cite{lodato10} have verified this numerically.
With $\alpha_1=0.3$ then we find the corresponding $\alpha_2=2.66$. We
note that varying $\alpha$ affects only the timescales of the effects
discussed in this work.

\subsection{Steady State Discs}

In this section let us consider steady state discs. We set $ \partial
\bm{L}/\partial t=0$ then take the scalar product of
equation~(\ref{maineq}) with $\bm{l}$ to find
\begin{equation}
  0=\frac{1}{R}\frac{\partial}{\partial R}
  \left[3R\frac{\partial}{\partial R}(\nu_1 L)-\frac{3}{2}\nu_1L\right]
\label{int}
\end{equation}
because $\bm{T}_{\rm tid}\bm{. l}=0$ (see Section~\ref{tt}). The tidal
torque is zero when it acts in the same direction as the angular
momentum of the disc at that radius.  Equation~(\ref{int}) has the
solution
\begin{equation}
\nu_1 \Sigma= A + B R^{-\frac{1}{2}},
\label{nusig2}
\end{equation}
where $A$ and $B$ are constants to be determined by the boundary
conditions.  The first term is the accretion disc term for which
$\nu_1 \Sigma=\,\rm const$ (far from the inner radius) and the second
term is the steady state decretion disc term which has $\nu_1 \Sigma
\propto R^{-\frac{1}{2}}$ \citep{pringle91} and zero mass accretion
rate through the disc.

The mass transfer rate through the disc is
\begin{equation}
\dot M=2\pi R \Sigma (-v_{\rm R}),
\label{mdot}
\end{equation}
where the radial velocity in the disc is
\begin{equation}
v_{\rm R}=-\frac{3}{\Sigma R^\frac{1}{2}}\frac{\partial}{\partial R}\left(\nu_1 \Sigma R^\frac{1}{2}\right)-\nu_2R\Bigl\lvert\frac{\partial \bm{l}}{\partial R}\Bigr\rvert ^2
\label{radvel}
\end{equation}
\citep{pringle81}.  The accretion rate is defined so that $\dot M>0$
implies accretion, the material moves inwards, and $\dot M<0$ implies
decretion, the material moves outwards. The general steady state
surface density in a disc is
\begin{equation} \nu \Sigma = \frac{\dot M}{3\pi} \left[1-\left(
      \frac{R_\star} {R}\right)^{\frac{1}{2}}\right] + \nu_{1\star}
  \Sigma_\star \left( \frac{R_\star}{R}\right)^{\frac{1}{2}},
\label{gen2} 
\end{equation}
where $\Sigma_\star$ and $\nu_\star$ are the surface density and
viscosity at $R=R_\star$ and the steady rate of mass transfer through
the disc is $\dot M$. The first term is the normal accretion disc term
with $\nu \Sigma=0$ at $R=R_\star$ and so zero torque at the inner
boundary.  The second term satisfies $v_{\rm R}=0$ when
$\partial\bm{l}/\partial R$ is zero there. With the additional second
term, we can vary the mass of the disc and the accretion rate through
the disc independently even with a steady state disc. For a Be star
disc that is truncated by the tides of a neutron star companion we
take only the second term, the steady state decretion disc term (see
Section~\ref{trunc} for more on the truncation of the disc).

\subsection{Surface Density}

The density of the disc is
\begin{equation}
\rho(R,z)=\rho_0 \left(\frac{R_\star}{R}\right)^n {\rm e}^{-\left(\frac{z-z_0}{H}\right)^2},
\label{rho}
\end{equation}
where $z_0$ is the height of the disc about the equatorial plane and
is a function of radius, $R$. For a flat disc $z_0(R)=0$
everywhere. By integrating the density over the disc height, $z$, we
find the surface density of the disc to be
\begin{equation}
\Sigma=\int_{-\infty}^{\infty}\rho \, dz= \Sigma_\star \left(\frac{R_\star}{R}\right)^{n-\frac{3}{2}},
\label{sigbe}
\end{equation}
with the constant
\begin{equation}
\Sigma_\star=\rho_0 \sqrt{\pi} H_\star .
\label{sigstar}
\end{equation}
\cite{wood97} find that at the stellar radius the disc aspect ratio is
$H_\star/R_\star \approx 0.04$. We choose a typical density of
$\rho_0= 10^{-11}\,\rm g\,cm^{-3}$ and $n=7/2$ and we find
\begin{align}
\frac{\Sigma_\star}{\rm g/cm^2}= &
\,0.49\left(\frac{H_\star/R_\star}{0.04}\right)
\left(\frac{R_\star}{10\,\rm R_\odot}\right)\left(\frac{\rho_0}{ 10^{-11}\,\rm g\,cm^{-2}}\right)\!\!.
\end{align}
We justify the choice of $n$ through the viscosity.

\subsection{Viscosity}
\label{iso}

For an isothermal disc the sound speed $c_{\rm s}=\,\rm const$ and so
the disc scale height is $H=c_{\rm s}/\Omega \propto R^{\frac{3}{2}}$.
We use the $\alpha$-prescription for the viscosities as in
equation~(\ref{viscs}) and find
\begin{equation}
\nu_i=\alpha_i \left(\frac{H}{R^{\frac{3}{2}}}\right)^2 (GM_1)^{\frac{1}{2}} R^{\frac{3}{2}},
\label{nu1}
\end{equation}
where $i=1,2$. With equation~(\ref{nusig2}) we have $\nu_1\Sigma\propto
R^{-\frac{1}{2}}$ in a steady state decretion disc and
with~(\ref{sigbe}) we found $\Sigma \propto R^{-n+\frac{3}{2}}$. The
viscosity must therefore be parametrised as
\begin{equation}
\nu_i=\nu_{i\star}\left(\frac{R}{R_\star}\right)^{n-2}.
\label{viscosityp}
\end{equation}
However, because $\nu_i \propto R^\frac{3}{2}$ with the
$\alpha$-prescription, for a consistent model we must choose
$n=7/2$. We find the viscosities at the stellar radius to be
\begin{align}
\nu_{i\star}= \,& 1.69\times 10^{17}\left(\frac{\alpha_i}{2.66}\right) \left(\frac{H_\star/R_\star}{0.04}\right)^2 \cr
& \times \left(\frac{M_1}{17\,\rm M_\odot}\right)^\frac{1}{2}\left(\frac{R_\star}{10\,\rm R_\odot}\right)^\frac{1}{2} \,\rm cm^2\,s^{-1}
\end{align}
for $i=1,2$. We note that if the disc was not isothermal, then we
would change the radius power law for the scale height, $H$, of the
disc. The value of $n$ decreases so that the surface density would
fall off more slowly with radius. This does not affect the shape of the
warped disc in the numerical models in Section~\ref{numerical}, only
the scale height and surface density of the disc.  \cite{porter03b}
found that with a smaller value of $n$ he was able to better reproduce
the observed continuum emission.

The first viscous timescale describes the timescale over which mass
flows through the disc and the second viscous timescale describes the
how the disc smoothes out the warp and twist. The viscous timescales
in the disc are
\begin{align}
t_{\nu_i}  =  \frac{R^2}{\nu_i}=  \frac{ R_\star^{2}}{\nu_{i\star}} \left(\frac{R}{R_\star}\right)^{\frac{1}{2}}
\end{align}
and so we find
\begin{align}
t_{\nu_i} = & \,\, 33.2\left( \frac{\alpha_i}{2.66}\right)^{-1}
\left(\frac{R_\star}{10\,\rm R_\odot}\right)^\frac{3}{2}
\left(\frac{M_1}{17\,\rm M_\odot}\right)^{-\frac{1}{2}}\cr
 & \times \left(\frac{H_\star/R_\star}{0.04}\right)^{-2}
\left(\frac{R}{R_\star}\right)^{\frac{1}{2}}\,\rm d
\label{tnu2}
\end{align}
for $i=1,2$.  The viscous timescales increase with radius in the
disc. For example, the minimum timescales at the inner edge of the
disc at the stellar radius are $t_{\nu_1}=294\,\rm d$ and
$t_{\nu_2}=33.2\,\rm d$.  An accretion disc reaches steady state on
the first viscous timescale.  The disc smoothes out on the second
viscous timescale.

\subsection{Tidal Truncation of the Disc}
\label{trunc}

Providing that the orbital period is shorter than the viscous
timescale, $t_{\nu_1}$, in the outer parts of the disc then the tidal
torques from the companion neutron star can truncate the disc.  In the
circular orbit case, the tidal truncation radius, $R_{\rm t}$, can be
estimated as where ballistic orbits begin to cross
\citep{paczynski77}. For example, in Figure~6 of \cite{martin08} this
radius is shown as a function of mass ratio of the binary. For Be star
systems, the mass ratio is in the approximate range $M_2/M_1=0.1-0.5$
so the truncation radius is $R_{\rm t}\lesssim 0.5a$, where $a$ is the
semi-major axis of the orbit. In the eccentric binary case, the disc
radius can be estimated by considering the effects of tidal
resonances. As the eccentricity of the binary increases, the
truncation radius of the disc decreases \citep{artymowicz94}.  The
neutron star orbit will be larger than the outer radius of the disc
unless the eccentricity is close to 1 or there is a large misalignment
angle.

We note that this has implications for theories of Type~II X-ray
outbursts. It has previously been suggested that they occur when the
Be star disc becomes sufficiently large that the neutron star passes
through the decretion disc. This is only possible if the disc does not
becomes truncated fast enough which may happen if the eccentricity is
close to 1 or if the misalignment angle is large enough that the
truncation is not efficient. \cite{okazaki07c} suggested that the
truncation becomes inefficient for large misalignments
$\theta>60^\circ$. In this work we consider smaller misalignments and
note that observed misalignments of Be X-ray binaries are relatively
small \citep[e.g.][]{MTP09}.

\section{Tidal Warping}
\label{tidal}

It is widely suggested that the warp and precession observed in Be
star discs may be caused by a misalignment between the spin axis of
the Be~star and the orbit of the binary companion.  Here we consider
further the tidal torque.

\subsection{Tidal Torque}
\label{tt}

The tidal torque in the frame of the binary orbit, of radius $R_{\rm
  b}$, is approximately
\begin{equation}
  \mathbf{T}_{\rm tid}=-\frac{GM_2R\Sigma }{2R_{\rm b}^2}
  \left[b_{3/2}^{(1)}\left(\frac{R}{R_{\rm b}}\right)\right]
  (\bm{e_z}.\bm{l})(\bm{e_z}\times \bm{l})
\label{tidaltorque}
\end{equation}
\citep{L00,OD01} to first order in the angle of tilt between the disc
and the orbital plane. Here $\bm{e_z}$ is a unit vector in the
direction of the binary orbital axis, $M_2$ is the mass of the binary
companion star and $R_{\rm b}$ is the binary separation. This tidal
torque is averaged over a binary orbit.  The Laplace coefficient of
celestial mechanics can be approximated for small $z$ by
\begin{equation}
  b_{3/2}^{(1)} \left(z\right) = 
  \frac{3}{2}z \left[ 1+ \frac{15}{8}z^2+\frac{175}{64}z^4+...\right].
\label{expansion}
\end{equation}
We note that it does indeed satisfy $\bm{l}.\bm{T}_{\rm tid}=0$ (see
Section~\ref{structure}) so that the tidal torque is zero when the
disc and orbit are aligned.  Thus even for large tilt angles the tidal
torque term is likely to have this form, although with modified
coefficients.

\subsection{Orbital Parameters}

Most Be star binaries have eccentric orbits and so the tidal torque
will not be constant over an orbit. In an eccentric orbit, the stars
spend a short time close to periastron but here, the torque will be
stronger than in a circular orbit. Because the orbital period of the
system is much smaller than the timescale on which the tidal torque
acts (see Section~\ref{applications}) we can average the torque over
an orbit. In this Section we find, for the average torque over an
eccentric orbit, the equivalent average binary separation.

With Kepler's third law the binary semi-major axis is
\begin{align}
a  =\left(\frac{GMP^2}{4\pi^2}\right)^{\frac{1}{3}},
\end{align}
where $P$ is the orbital period and $M=M_1+M_2$ and so
\begin{align}
 a = 6.50\times 10^{12}\left(\frac{M}{18.4\,\rm M_\odot}\right)^\frac{1}{3}
\left(\frac{P}{24.3\,\rm d}\right)^\frac{2}{3}\,\rm cm.
\label{semimajor}
\end{align}
From Kepler's first law, the separation of the binary stars in an
eccentric orbit satisfies
\begin{equation}
R_{\rm b}=\frac{a(1-e^2)}{1+e\cos \phi},
\label{firstlaw}
\end{equation}
where $e$ is the eccentricity of the orbit and $\phi$ is the azimuthal
angle around the orbit. With Kepler's second law
\begin{equation}
\frac{d}{dt}\left(\frac{1}{2}R_{\rm b}^2\dot \phi\right)=0,
\label{secondlaw}
\end{equation}
where $\dot \phi=d\phi/dt$. We integrate this equation twice over an orbit
to find
\begin{equation}
\frac{1}{2}R_{\rm b}^2\dot \phi=\frac{\pi a^2}{P}(1-e^2)^\frac{1}{2}.
\label{thetadot}
\end{equation}
The tidal torque found by \cite{OD01} in equation~(\ref{tidaltorque})
has a factor of $1/R_{\rm b}^3$ which we must replace with its average
in time
\begin{align}
 \left<\frac{1}{R_{\rm b}^3}\right> & =\frac{1}{P}\int_0^P\frac{1}{R_{\rm b}^3}\, dt \cr
 & =\frac{1}{a^3( 1-e^2)^\frac{3}{2}}.
\end{align}
The average separation is thus
\begin{equation}
\bar R_{\rm b}= a(1-e^2)^\frac{1}{2}
\label{r3}
\end{equation}
\citep[see e.g.][]{holman97}.  In Section~\ref{applications} we find
the tidal timescale for observed Be star binaries and there we must use
equation~(\ref{tidaltorque}) replacing $R_{\rm b}$ with $\bar R_{\rm
  b}$ so that we do not underestimate the strength of the tidal torque
in an eccentric system.

\subsection{Tidal Warp Radius}
\label{tidwarp}

The tidal torque balances the viscous torque in the disc at the tidal
warp radius. Outside of the tidal warp radius the disc is dominated by
the tidal torque and inside of this radius by the viscous torques. The
tidal warp radius is
\begin{equation}
  R_{\rm tid}=\left[\frac{2 \nu_{2\star}(GM_{\rm
        1})^{\frac{1}{2}}{\bar R_{\rm b}^3}}
    {3 GM_2R_\star^{n-2}}\right]^{\frac{2}{11-2n}}
  \label{rtid2}
\end{equation}
\citep{MPT07,Martintid}. Note that in deriving this radius, only the
first term in the expansion in equation~(\ref{expansion}) has been
used. This is the same radius for accretion and decretion discs which
have only a tidal torque because it is independent of surface
density. We expect that well outside this radius the binary torque
term, coupled with the viscosity, flattens the disc and it aligns with
the binary orbital plane. If the disc does not extend up to the tidal
radius, the torque can still have an effect on the disc and cause the
disc to move towards alignment with the binary orbit even if they
don't completely align.

With the standard parameters used in the previous Section we find with
$n=7/2$ that
\begin{align}
  R_{\rm tid}  = & \,3.68\times 10^{12}\left(1-e^2\right)^\frac{3}{4}
\left(\frac{P}{24.3\,\rm d}\right)
\left(\frac{\alpha_2}{2.66}\right)^\frac{1}{2}\cr
& \times 
\left(\frac{M_1}{17\,\rm M_\odot}\right)^\frac{1}{2}
\left(\frac{M_2}{1.4\,\rm M_\odot}\right)^{-\frac{1}{2}}
\left(\frac{M}{18.4\,\rm M_\odot}\right)^\frac{1}{2}\cr
& \times 
\left(\frac{R_\star}{10 \,R_\odot}\right)^{-\frac{1}{2}}
\left(\frac{H_\star/R_\star}{0.04}\right)
 {\,\rm cm}.
\label{rtid}
\end{align}
If $n=4$ (as we have used for our numerical models in
Section~\ref{numerical}) then we find $R_{\rm tid}=6.1\times
10^{12}\,\rm cm$ for these general parameters. With the small orbital
period shown here the warp radius is within the disc. However for Be
star systems of longer period, it may be outside of the disc. In this
case the viscous torques in the disc always dominate the tidal
torques. The disc is relatively flat, but may be tilted from the
equator of the Be star and precessing.

\subsection{Tidal Timescale}

The timescale on which the tidal torque acts is equal to the viscous
timescale at $R=R_{\rm tid}$ by definition. It scales as
\begin{align}
t_{\rm tid}  = \frac{R_\star^2}{\nu_{2\star}}\left(\frac{R_{\rm tid}}{R}\right)^\frac{3}{2}.
\label{tidtime}
\end{align}
We parametrise this as
\begin{equation}
t_{\rm tid}=t_{\rm tid\star} \left(\frac{R}{R_\star}\right)^{-\frac{3}{2}},
\end{equation}
where the tidal timescale at $R=R_\star$ is
\begin{align}
t_{\rm tid\star} = &\,\, 1.1
\left(1-e^2\right)^\frac{9}{8}
\left(\frac{P}{24.3\,\rm d}\right)^\frac{3}{2}
\left(\frac{\alpha_2}{2.66}\right)^{-\frac{1}{4}}
\cr & \times
\left(\frac{M_1}{17\,\rm M_\odot}\right)^\frac{1}{4}
\left(\frac{M_2}{1.4\,\rm M_\odot}\right)^{-\frac{3}{4}}
\left(\frac{M}{18.4\,\rm M_\odot}\right)^{\frac{3}{4}}
\cr & \times
\left(\frac{R_\star}{10 \,R_\odot}\right)^{-\frac{3}{4}}
\left(\frac{H_\star/R_\star}{0.04}\right)^{-\frac{1}{2}}
\,\rm yr.
\label{tidtime2}
\end{align}
The disc aligns with the binary orbital plane on this timescale.  This
is longer than the orbital period and so the disc inclination should
eventually reach a steady state.

\section{Evolution of the Misaligned Disc}
\label{numerical}

We solve equation~(\ref{maineq}) numerically with a first order
explicit method \citep{pringle92} including the non-linear terms. We
concentrate on the effects of the tidal torque, without which the disc
would remain flat and aligned with the spin axis of the Be star. The
size of Be star discs has been observed in a few cases.
\cite{grundstrom06} used results from a H$\alpha$ monitoring campaign
to find the Be~stars that they considered had discs with an outer edge
in the range $3-11\, R_\star$.  In this work we choose the Be star to
have an outer truncation radius of $R_{\rm out}=10\,R_\star$.  We use
200 grid points distributed logarithmically between the inner radius
$R_{\rm in}=R_\star$ and the outer radius $R_{\rm out}$.

The inclination of the disc angular momentum to the binary orbital
axis is
\begin{equation}
i(R)=\cos^{-1}(l_z)
\end{equation} 
and the azimuthal angle is
\begin{equation}
\omega(R)=\tan^{-1}\left(\frac{l_y}{l_x}\right).
\end{equation}
The tidal precession can be parametrised as 
\begin{equation}
\bm{T}_{\rm tid}=\bm{\Omega}_{\rm tid}\bm{\times}\bm{L}
\end{equation}
\citep[see][]{Martintid}, where we choose coordinates so that
\begin{equation}
\bm{\Omega}_{\rm tid}=\frac{1}{t_{\nu_2}(R_{\rm tid})}\left(\frac{R}{R_{\rm tid}}\right)^\frac{3}{2}\bm{e}_z.
\end{equation}
This term has the effect of precessing the disc about the $\bm{e}_z$
axis when the disc in not aligned with the binary orbital plane. This
corresponds to the first term in the expansion given in
equation~(\ref{expansion}). 

Initially the disc has very little mass that is flat and aligned with
the spin of the Be star as given in equation~(\ref{jadd}), where
$\theta$ is the misalignment between the spin of the Be star and the
binary orbital axis.  Mass is injected on to the disc at a constant
rate $\dot M$ at a radius of $R_{\rm add}=1.1\, R_\star$. The mass is
always added with an angular momentum aligned with the spin axis of
the Be star which in all models is misaligned to the binary orbital
axis by $\theta=25^\circ$.

We need three boundary conditions at the inner and outer edges of the
disc. We choose two conditions on the direction of the angular
momentum at both edges, $\partial \bm{l}/\partial R=0$ with
$|\bm{l}|=1$. We need one more condition on the surface density. In
this work we consider three different surface density inner boundary
conditions that we describe in the Section~\ref{inner}.  We use a zero
radial velocity surface density outer boundary condition, $v_{\rm
  R}=0$, at $R_{\rm out}$ because we assume that the outer radius is
truncated by the tidal torque from the neutron star. Tidal torques
truncate the edge of a disc over a small radius
\citep[e.g.][]{ichikawa94,martin11}.  \cite{okazaki02} use a 3D SPH
code to model a Be star disc that is coplanar with the neutron star
orbit. They find that the disc is truncated if $\alpha_1 \ll 1$. With
the neutron star companion, the disc becomes denser than the disc
around an isolated Be star. \cite{okazaki07b} extend this model to a
disc that is inclined to the binary orbital plane.  They have material
accreted on to the Be star disc for 50 orbital periods and then they
turn off the accretion. They find that there is little precession
until the accretion is turned off. With a one-dimensional code we can
evolve the disc for a longer period of time.

We choose the first viscosity
\begin{equation}
\nu_1=0.1\left(\frac{R}{R_\star}\right)^2\, R_\star^2 t_{\rm c}^{-1}
\end{equation}
and the second viscosity 
\begin{equation}
\nu_2=\left(\frac{R}{R_\star}\right)^2 \, R_\star^2 t_{\rm c}^{-1},
\end{equation}
where we have defined the time unit
\begin{equation}
t_{\rm c}=t_{\nu_2}(R_\star),
\label{tc}
\end{equation}
which with equation~(\ref{tnu2}) for our standard parameters
$t_{\nu_2}=58.8\,\rm d$. With these viscosities the viscous timescales
are constant in radius through the disc and the evolution proceeds at
the same rate at all radii. This is slightly different from the
prescription we found in Section~\ref{structure}.

\subsection{Inner Boundary Condition}
\label{inner}

The inner boundary condition is an approximation to a torque applied
close to the star that depends on the magnetic field of the Be
star. We assume that the magnetic axis is aligned with the spin of the
Be star. We first consider a torque that is strong enough to prevent
all accretion back on to the Be star. This is equivalent to $v_{\rm
  R}=0$ at $R_\star$ or else
\begin{equation}
\frac{\partial}{\partial R}\left( \nu_1 \Sigma
R^\frac{1}{2}\right)\Bigl\lvert_{R=R_\star}=0
\label{firstbc}
\end{equation}
\citep{pringle91}.  Next we consider a disc with no additional torque
from the star so that the inner boundary condition is $\Sigma=0$ at
$R=R_\star$ \citep{pringle81}. Finally we choose a magnetic torque
that varies with the inclination of the inner edge of the disc. The
torque is strongest when the disc is aligned with the equator of the
star but weakens as the inner disc moves away from the equator. We
parametrise the inner boundary condition with
\begin{equation}
\frac{\partial}{\partial R}\left( \nu_1 \Sigma
R^\frac{1}{2}\right)\Bigl\lvert_{R=R_\star}=K\frac{\nu_{1\star}\Sigma_\star}{R_\star^\frac{1}{2}}
\left[\frac{\Omega_\star/\Omega_{\rm star}}{\cos(i_\star-\theta)}-1\right],
\label{secondbc}
\end{equation}
where $\Omega_\star=\Omega(R_\star)$, $i_\star=i(R_\star)$,
$\Omega_{\rm star}$ is the angular velocity of the Be star and $K$ is
a constant we can vary. We note that the zero radial velocity and zero
surface density inner boundary conditions represent the two limiting
cases of possible inner boundary conditions. We consider all these
boundary conditions further in the Appendix and show that the
evolution is the same with an extra torque term acting on the disc.

\subsection{Zero Radial Velocity Inner Boundary}
\label{zerovr}

 The zero radial velocity inner boundary condition is given in in
 equation~(\ref{firstbc}). In this case flow is prevented from both
 the inner and outer edge of the disc and so there is no mass loss
 from the disc. Mass is added to the disc at a constant rate and so
 the mass of the disc increases linearly in time
\begin{equation}
M_{\rm d}= \left(\frac{t}{t_{\rm c}}\right)\, \dot M t_{\rm c}.
\end{equation}
For example if the mass accretion rate on to the disc is $\dot
M=10^{-9}\,\rm M_\odot\,yr^{-1}$ and the second viscous timescale at
the inner edge of the disc is $t_{\nu_2}(R_\star)=58.8{\,\rm d}=t_{\rm
  c}$. At a time of $t=243\,t_{\rm c}$ the mass of the disc would be
$M_{\rm d}=3.9\times 10^{-8}\,\rm M_\odot$.

\begin{figure*}
\begin{center}
  \epsfxsize=5.5cm \epsfbox{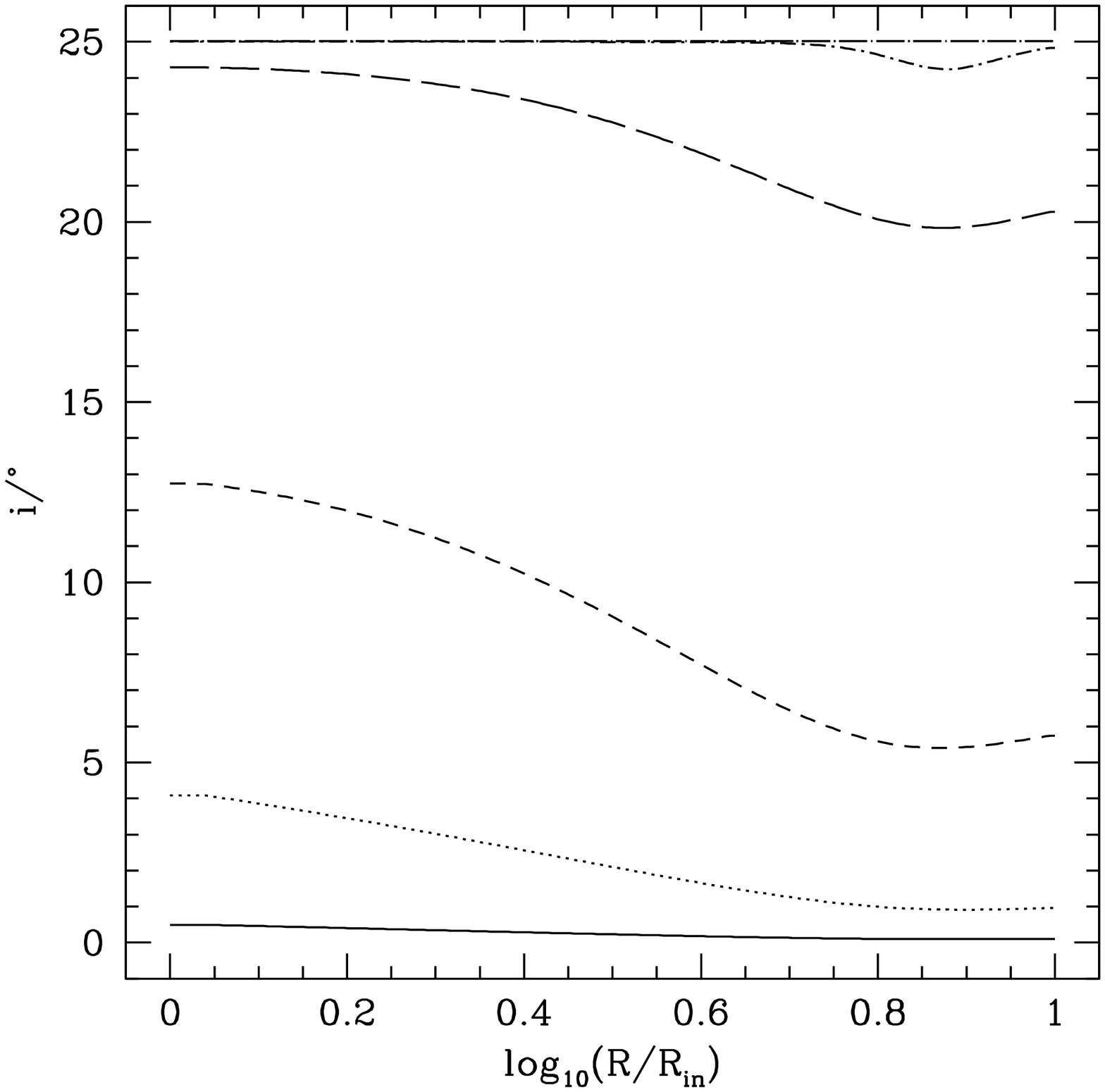}
  \epsfxsize=5.5cm \epsfbox{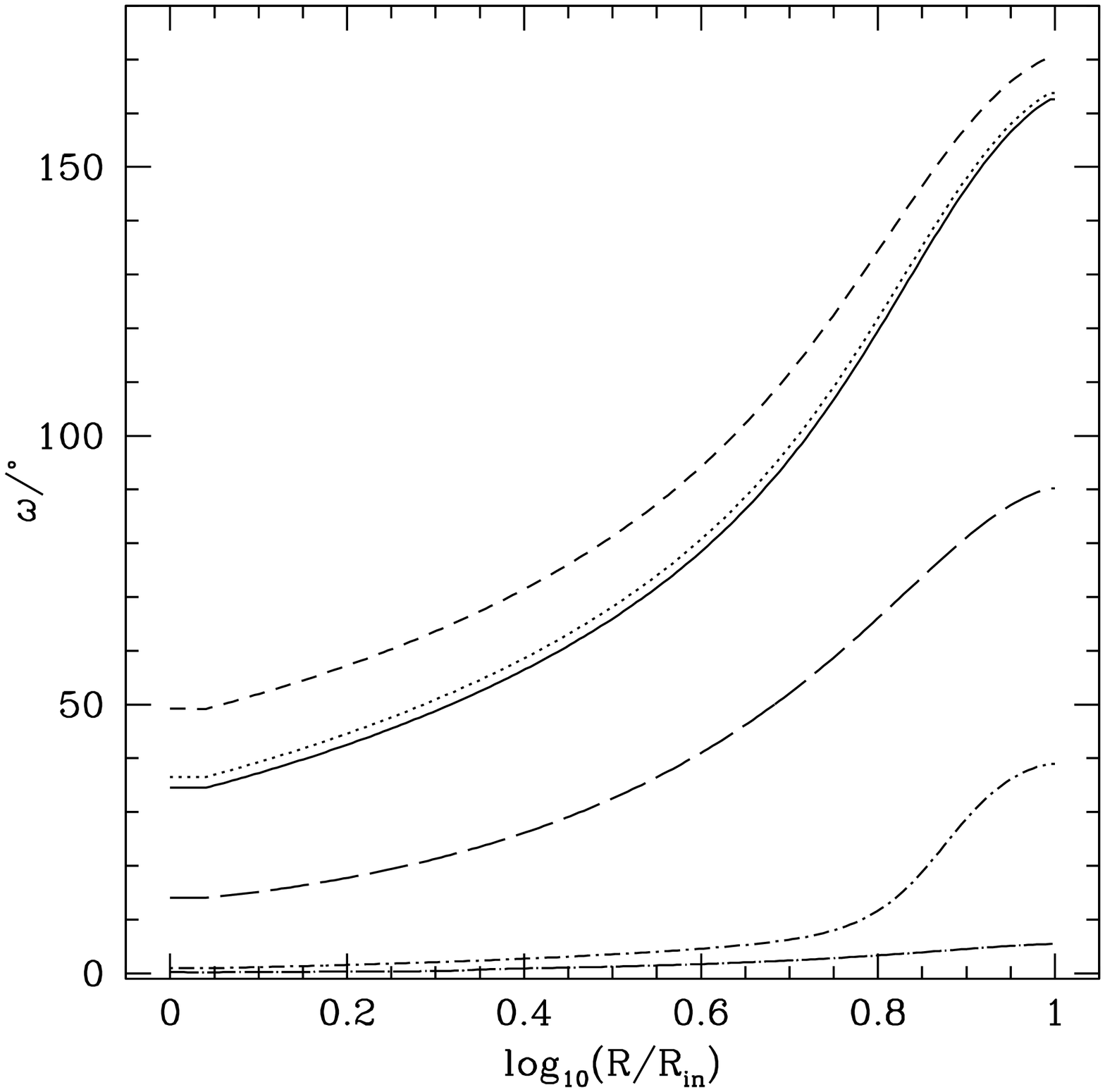}
  \epsfxsize=5.5cm \epsfbox{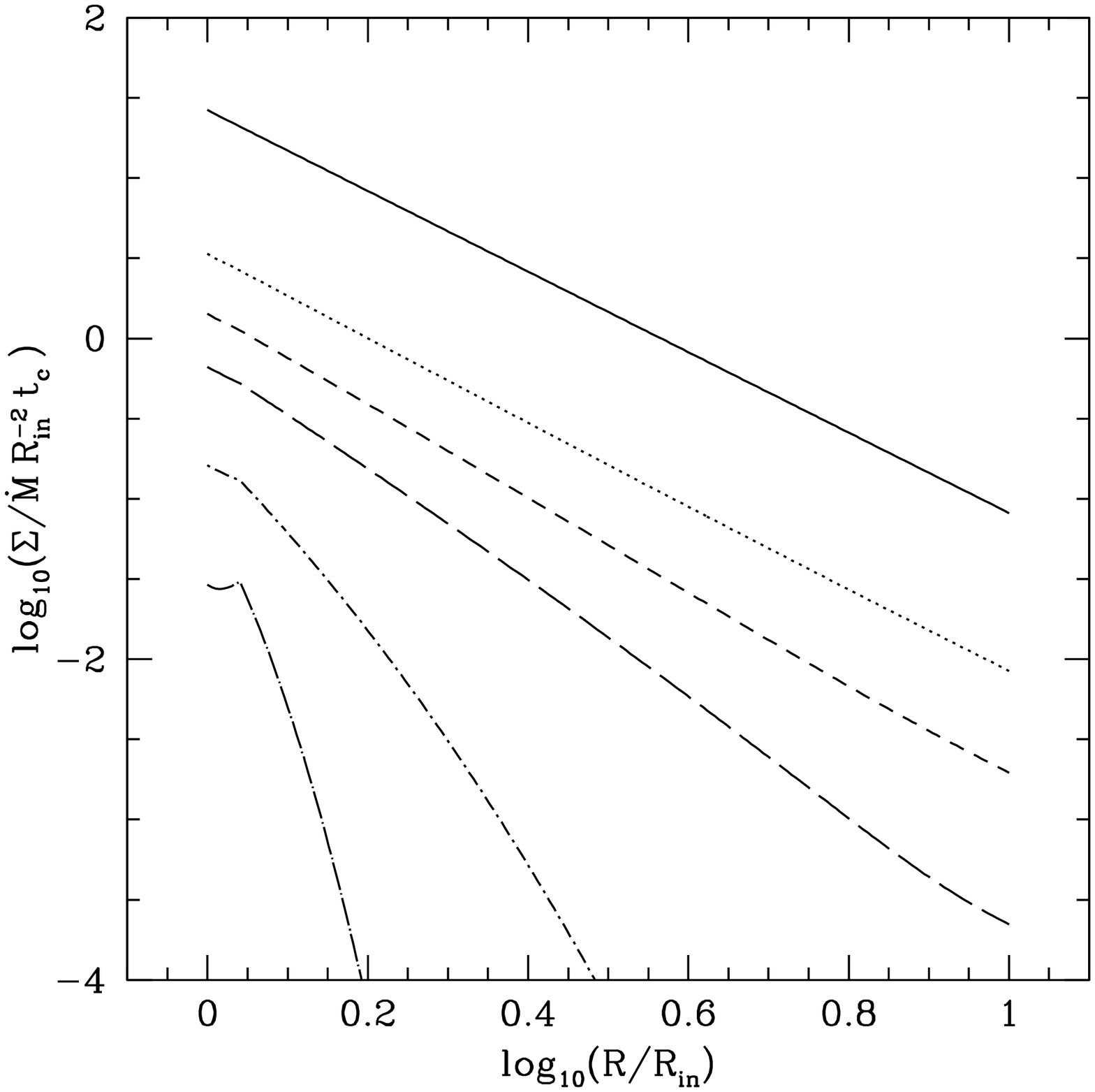}
  \caption[] {The disc has a warp radius of $R_{\rm tid}=5\,R_\star$
    and a misalignment of $\theta=25^\circ$. The inner boundary
    condition is $v_{\rm R}=0$. The evolution of the inclination of
    the disc angular momentum to the binary orbital axis (left), the
    azimuthal angle (middle) and the surface density (right) of the
    disc. The times shown are $t=0.037$ (long dot dashed lines),
    $0.33$ (short dot dashed lines), $3.0$ (long dashed lines), $9.0$
    (short dashed lines), $27.0$ (dotted lines) and $243 \,t_{\rm c}$
    (solid lines), where the time unit is defined in
    equation~(\ref{tc}). The mass of the disc increases linearly in
    time.}
\label{inc}
\end{center}
\end{figure*}

In Fig~\ref{inc} we show the inclination and azimuthal angle of the
angular momentum vector and the surface density of such a disc which
has a binary companion misaligned by $\theta=25^\circ$ and a tidal
warp radius of $R_{\rm tid}=5\,\rm R_\star$.  Mass is injected at
$R_{\rm add}$ and initially the mass moves outwards. Once it reaches
the outer boundary where the disc becomes truncated it reaches a
steady state where there is no flow in the disc. The surface density
continues to increase steadily in time once it has reached its steady
state of $\nu \Sigma \propto R^{-\frac{1}{2}}$ \citep{pringle91}. The
tidal torque is strongest at the outside of the disc and so initially
the outer parts of the disc align with the binary orbital axis and the
inner parts follow. The tidal timescale at the inner radius is $t_{\rm
  tid}(R_\star)=11.1 \,t_{\rm c}$ (with equation~\ref{tidtime}). At
this time, the disc has become significantly misaligned with the spin
of the Be star (left hand plot) and significantly twisted (middle
plot). The surface density has reached its quasi steady state. The
tidal timescale at the outer radius in the disc is $t_{\rm
  tid}(10\,R_\star)=0.35 \,t_{\rm c}$. At this time, the disc is still
close to alignment with the equator of the Be star.

\begin{figure*}
\begin{center}
  \epsfxsize=8.4cm \epsfbox{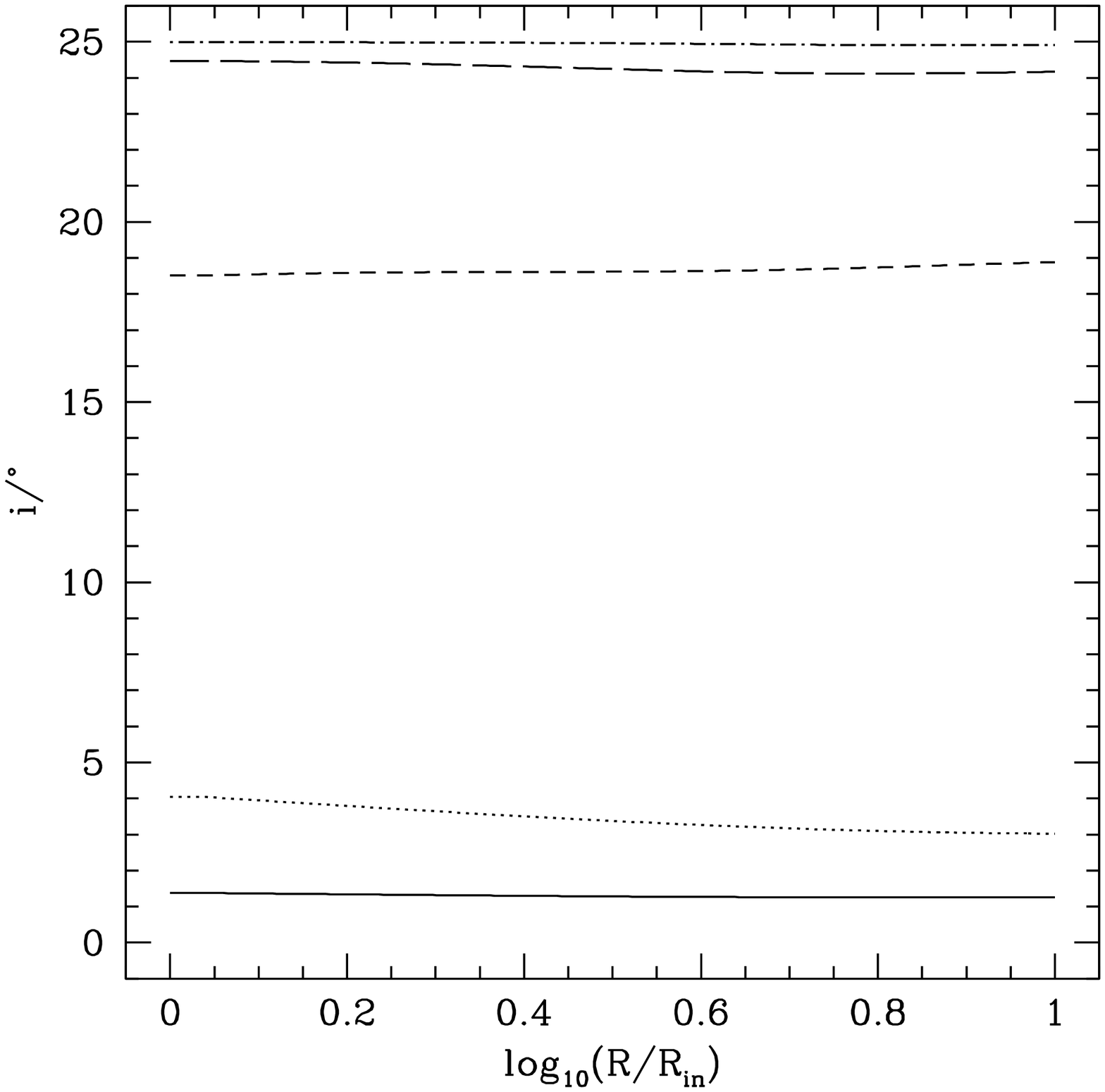}
  \epsfxsize=8.4cm \epsfbox{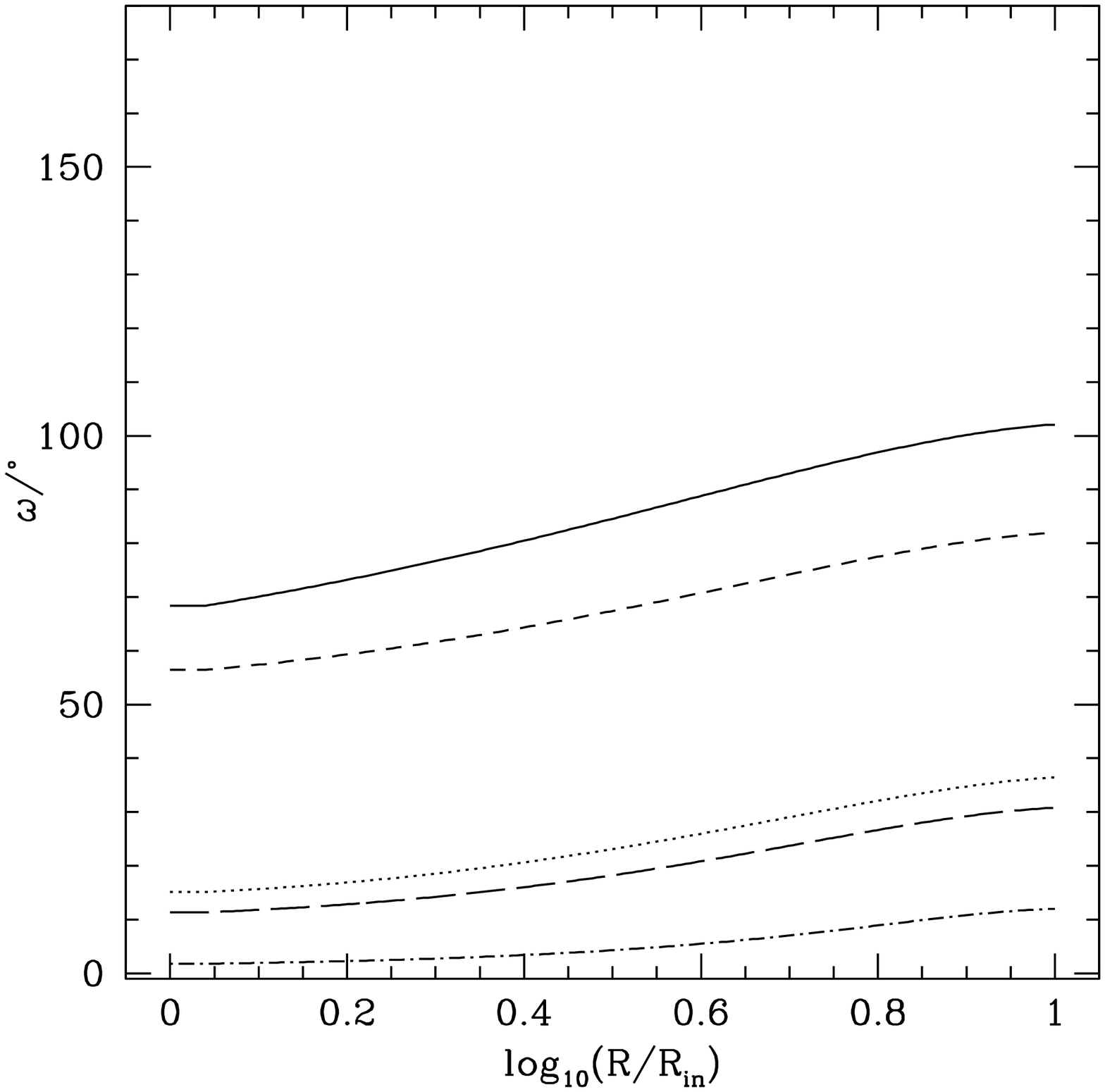}
  \epsfxsize=8.4cm \epsfbox{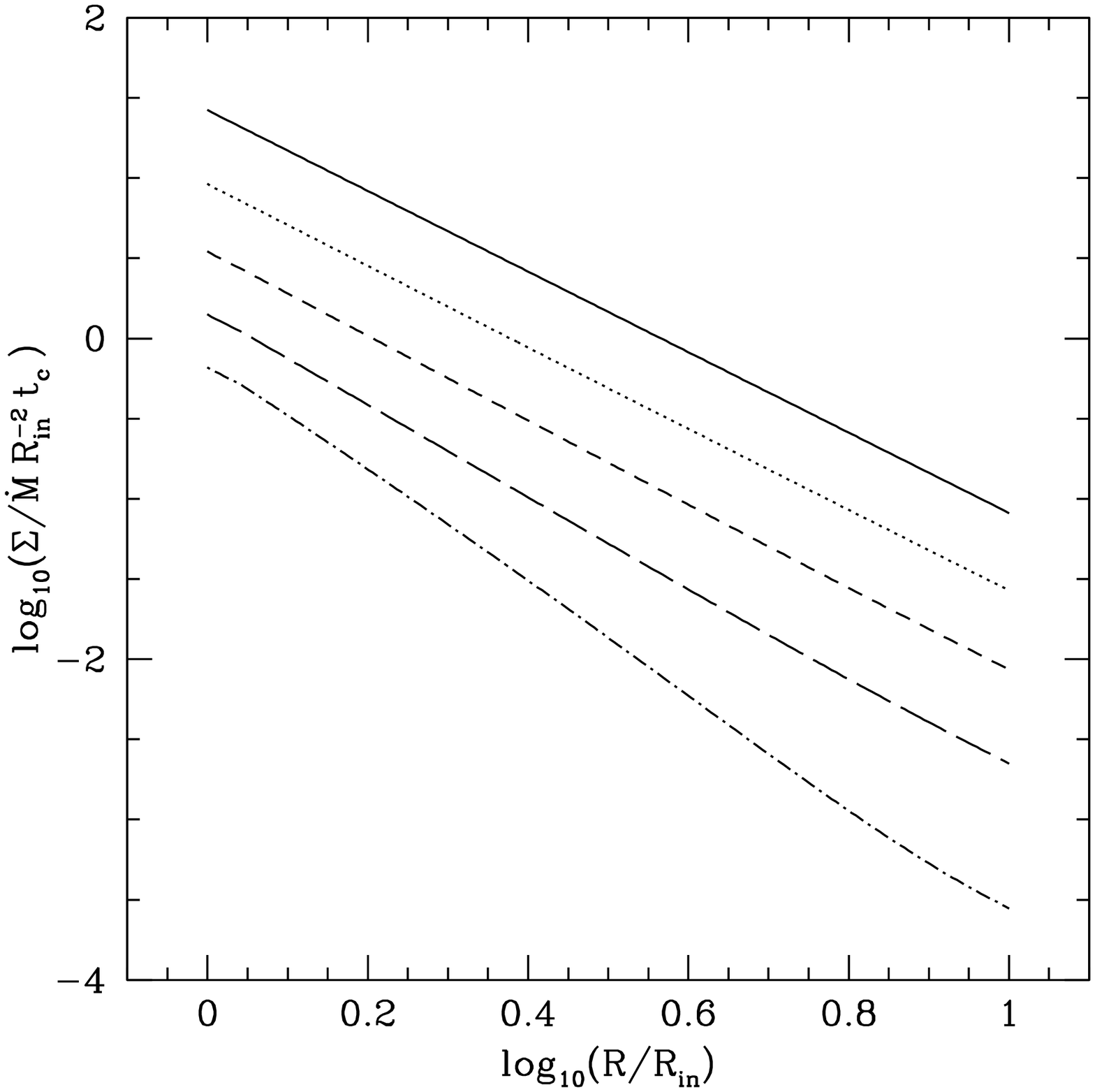}
  \epsfxsize=8.4cm \epsfbox{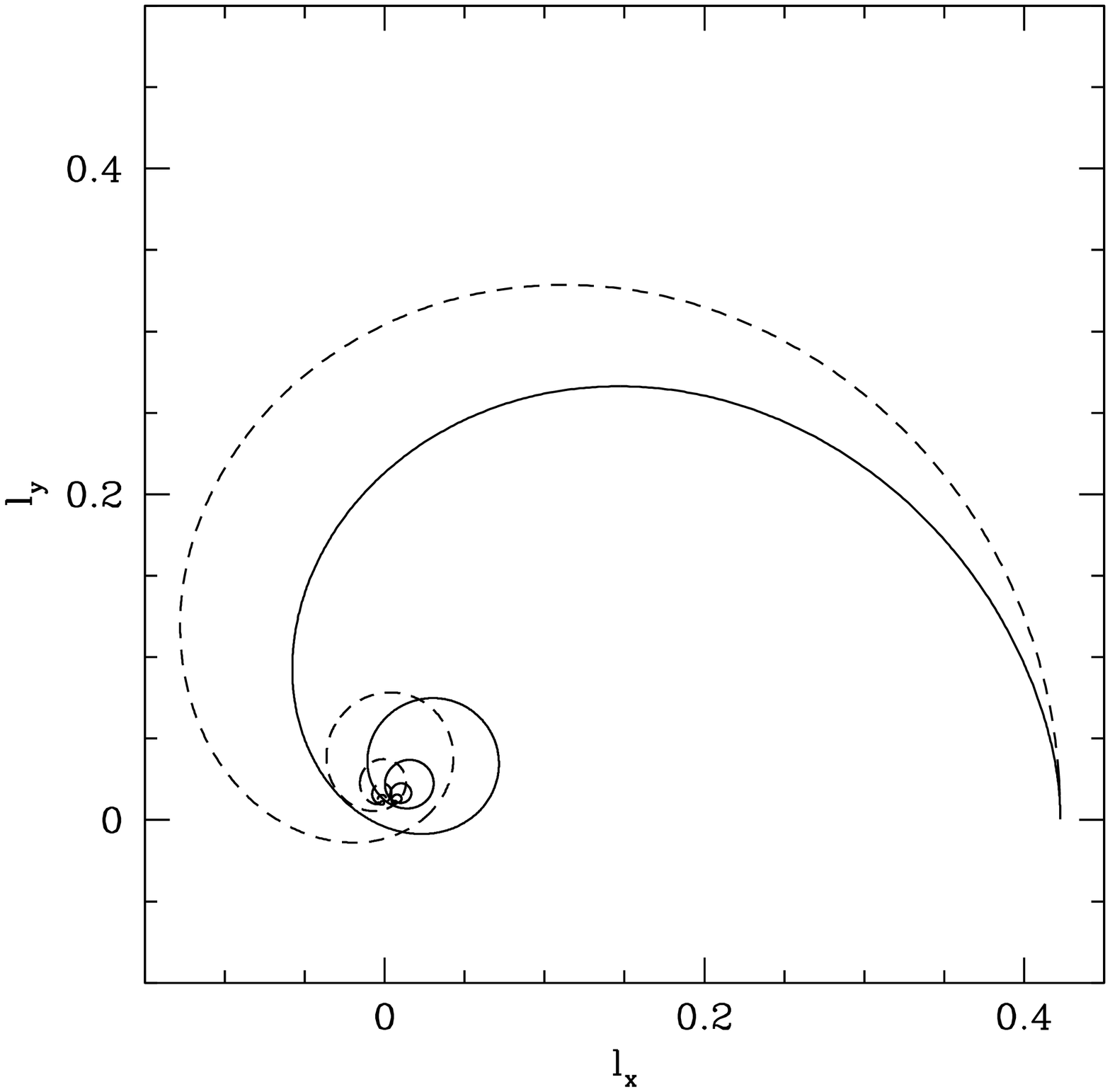}
  \caption[] {The zero radial velocity inner boundary condition for a
    disc with $R_{\rm tid}=20\,R_\star$ and $\theta=25^\circ$. The
    evolution of the inclination of the disc angular momentum to the
    binary orbital axis (top left), the azimuthal angle (top right)
    and the surface density (bottom left) of the disc in time. The
    times shown are $t= 3.0$ (dot-dashed lines), $9.0$ (long dashed
    lines), $27.0$ (short dashed lines) $81$ (dotted lines) and $243
    \,t_{\rm c}$ (solid lines), where the time unit is defined in
    equation~(\ref{tc}). The mass of the disc increases linearly in
    time. The bottom right plot shows the evolution of the angular
    momentum vector of the inner (solid line) and outer (dashed line)
    edges of the disc. At $t=0$ the disc is at $l_y=0$ on the right
    hand side of the plot.}
\label{inc2}
\end{center}
\end{figure*}

In Fig.~\ref{inc2} we consider a disc with a larger tidal warp radius
of $R_{\rm tid}=20\,R_\star$. This is outside of the outer edge of the
disc. The viscous torques in the disc are always larger than the tidal
torques.  The surface density evolution is the same as in
Fig.~\ref{inc} because it is independent of the tidal warp radius. The
disc is less warped and twisted with a larger tidal warp radius. It
moves more like a solid body. The tidal timescale at the inner edge of
the disc is $t_{\rm tid}(R_\star)=89.4 \,t_{\rm c}$. At this time the
disc has moved significantly from the initial alignment with the Be
star equator.  In the bottom right hand plot we show the evolution of
the angular momentum vectors at the inner and outer edge of the disc
in time. Initially the disc has $l_y=0$ and both vectors are the same,
on the right hand side of the plot. The disc precesses about the
binary orbital axis at the origin.

From the two figures in this Section we conclude that the Be star disc
aligns and precesses with the binary orbital plane on the tidal
timescale at the inner edge of the disc.  Once the disc mass is large
enough, the additional torque on the disc from the injected matter
becomes small compared with the torque from the binary companion in
the larger mass disc so the disc always becomes aligned with the plane
of the binary orbital axis, no matter what the warp radius is.

\subsection{Zero Torque  Inner Boundary Condition}
\label{full}

The second inner boundary condition that we consider is a zero torque
inner boundary condition normally associated with accretion discs 
\begin{equation}
\Sigma(R_\star)=\Sigma_\star=0.
\end{equation}
As before material is continuously injected at $R_{\rm add}$ and it
moves both outwards and inwards.  In $R>R_{\rm add}$ the disc behaves
as a decretion disc and in $R<R_{\rm add}$ as a normal accretion disc.

\begin{figure*}
\begin{center}
  \epsfxsize=5.5cm \epsfbox{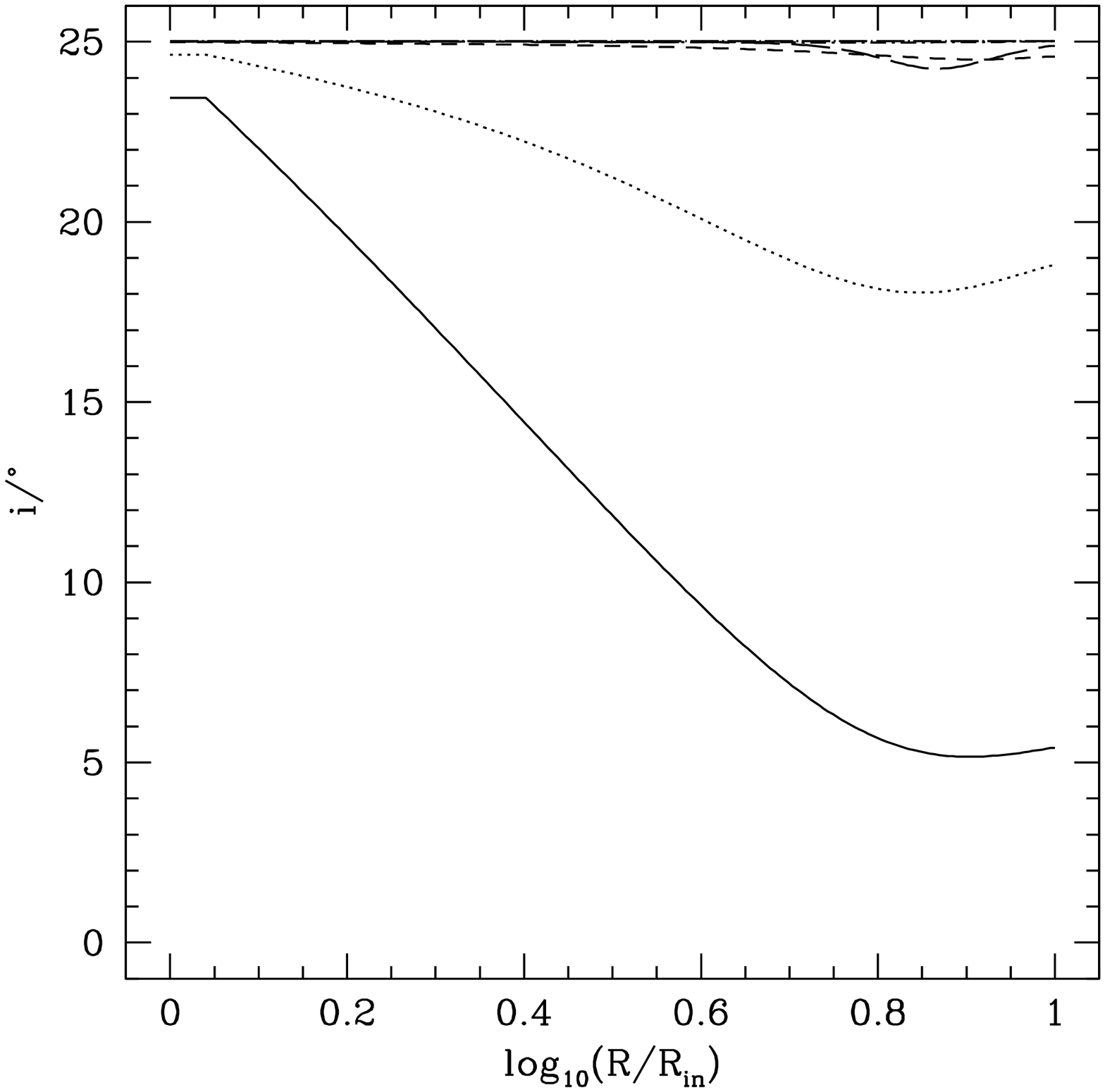}
  \epsfxsize=5.5cm \epsfbox{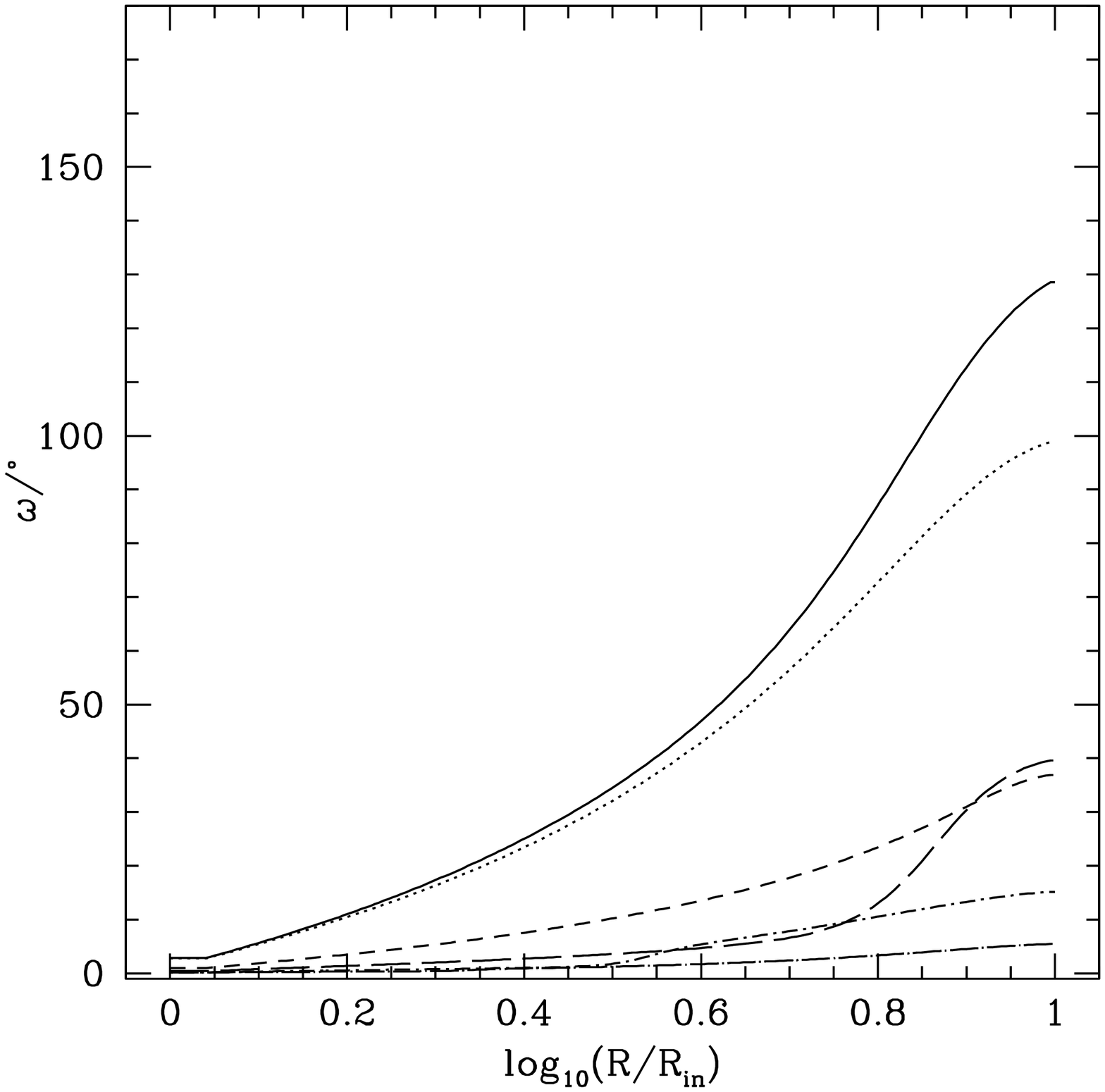}
  \epsfxsize=5.5cm \epsfbox{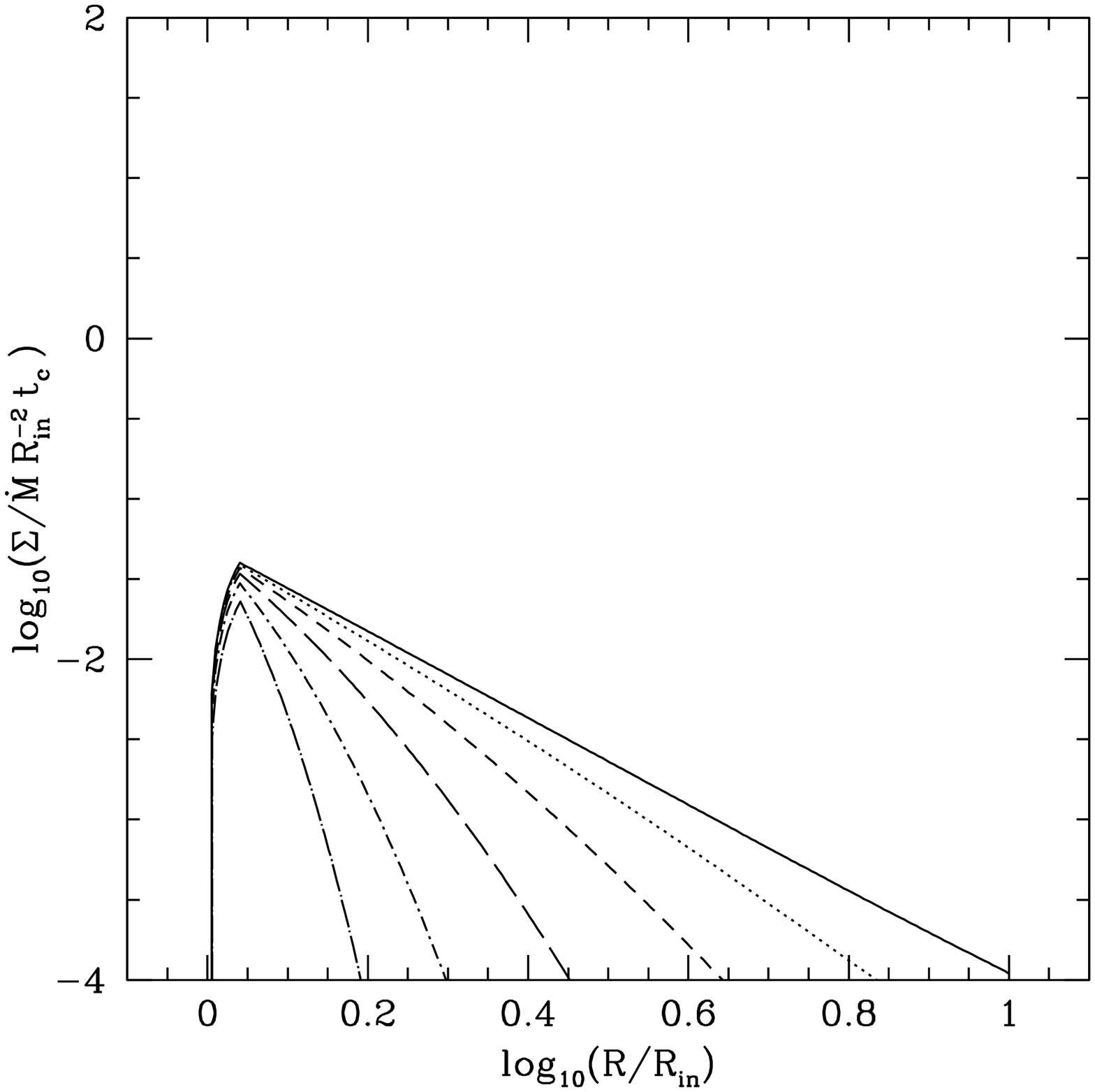}
  \caption[] {The inner boundary condition is $\Sigma_\star=0$ for a
    disc with $R_{\rm tid}=5\,R_\star$ and $\theta=25^\circ$. The
    evolution of the inclination (left), azimuthal angle (middle) and
    surface density (right) of the disc in time. The times shown are
    $t=0.037$ (long dot dashed lines), $0.11$ (short dot dashed lines),
    $0.33$ (long dashed lines), $1.0$ (short dashed lines), $3.0$ (dotted
    lines) and $81\, t_{\rm c}$ (solid lines). The time unit is defined in
    equation~(\ref{tc}).}
\label{acc5}
\end{center}
\end{figure*}

In Fig.~\ref{acc5} we plot the evolution of a system with $R_{\rm
  tid}=5\,R_\star$.  In this example the disc mass reaches a steady
state of $M_{\rm d}=0.36\,\dot M\,t_{\rm c}$. Because the disc mass is
lower than in the previous case, the mass addition torque is
relatively stronger compared with the tidal torque and so the inner
parts of the disc remain close to alignment with the Be star equator
while the outer parts tend towards alignment with the binary orbital
plane. We do not show the evolution of the angular momentum vector in
this Section because there is very little precession with the zero
torque angular momentum inner boundary condition.

\subsection{Varying Radial Velocity at the Inner edge}
\label{accretion}

\begin{figure*}
\begin{center}
  \epsfxsize=5.5cm \epsfbox{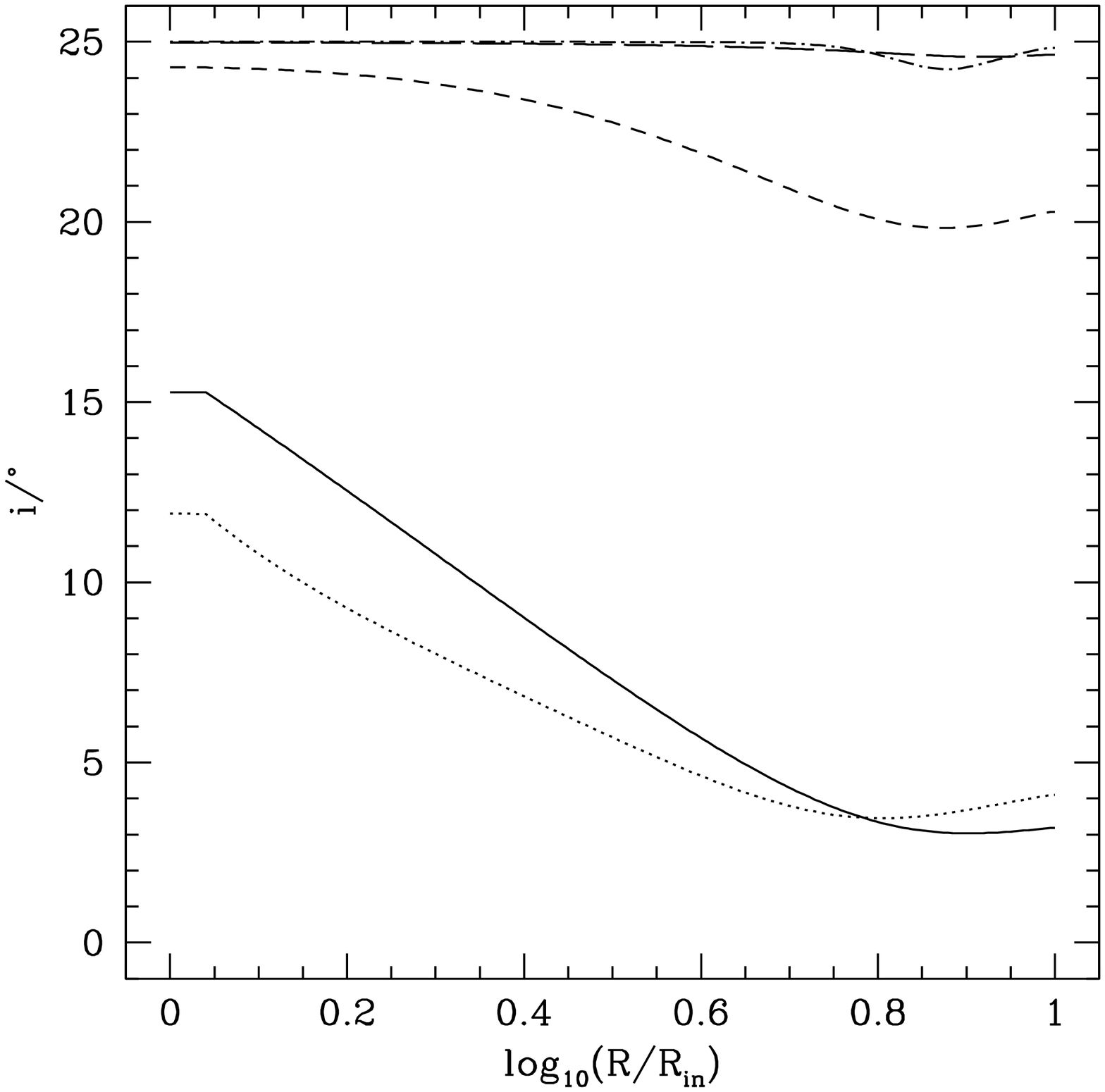}
  \epsfxsize=5.5cm \epsfbox{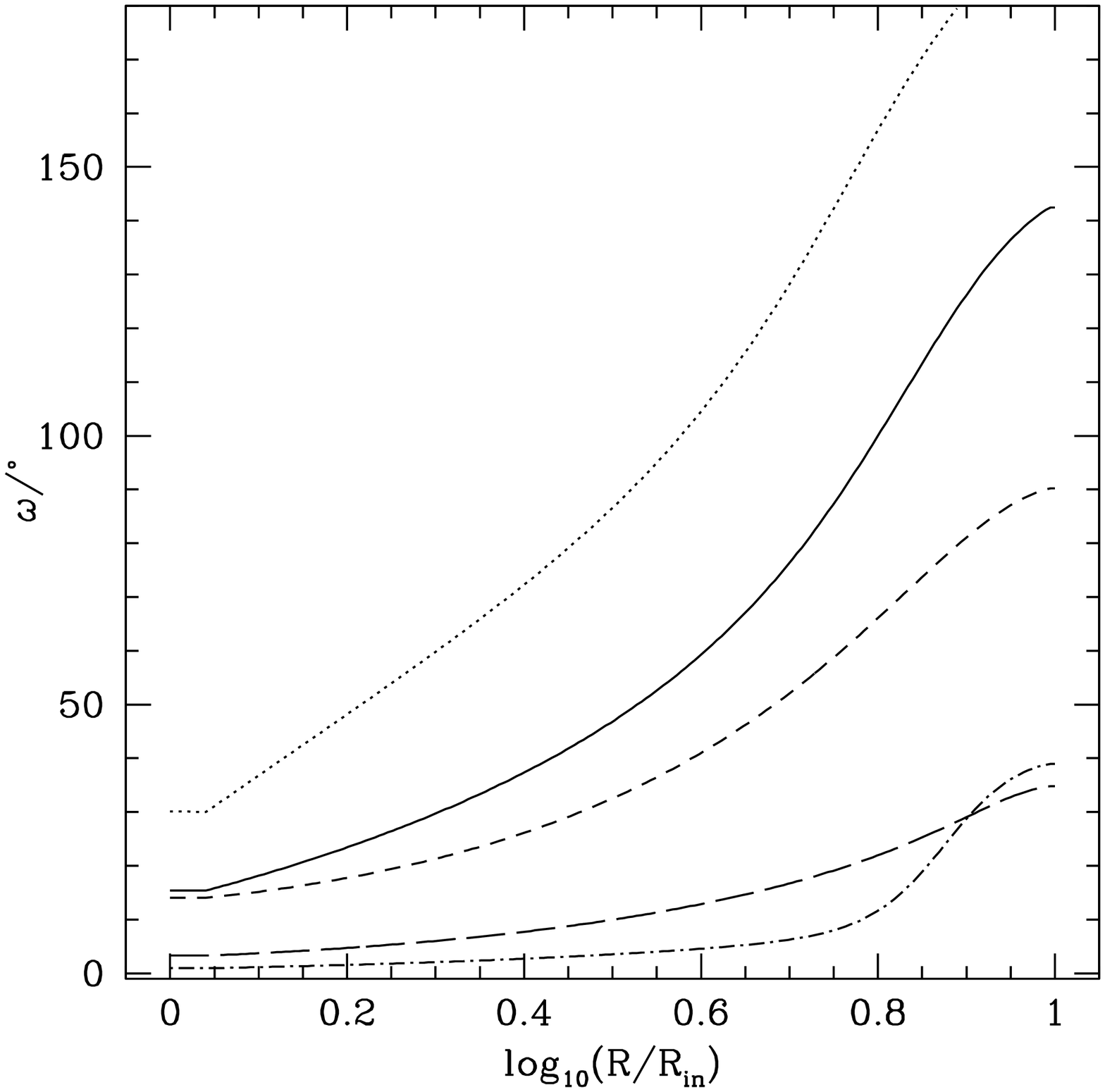}
  \epsfxsize=5.5cm \epsfbox{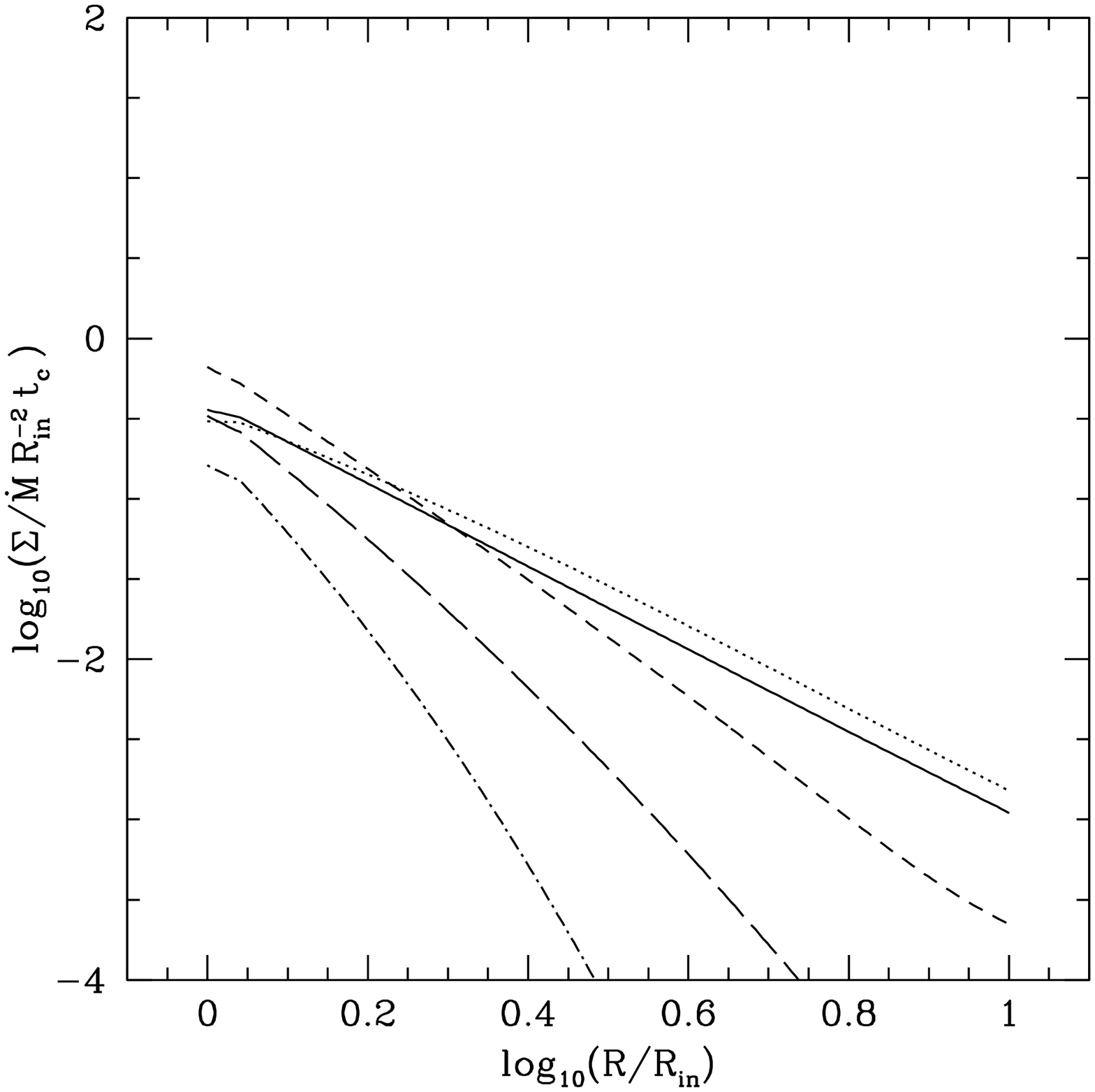}
  \caption[] {The inner boundary condition is the one given in
    equation~(\ref{secondbc}) that varies with inclination of the
    inner disc to to the Be star with $K=100$. The disc has $R_{\rm
      tid}=5\,R_\star$ and $\theta=25^\circ$. The evolution of the
    inclination (left), azimuthal angle (middle) and surface density
    (right) of the disc at times $t=0.33$ (dot dashed lines), $1.0$
    (long dashed lines), $3.0$ (short dashed lines), $9.0$ (dotted
    lines) and $27.0\, t_{\rm c}$ (solid lines) where the time unit is defined in
    equation~(\ref{tc}).}
\label{ss5}
\end{center}
\end{figure*}

In this Section we consider the inner boundary condition that varies
with the inclination of the inner disc to the equator of the Be star
given in equation~(\ref{secondbc}). Unless otherwise stated we
consider the Be star to be rotating at break up velocity so that
$\Omega_{\rm star}=\Omega_\star$.  In Fig.~\ref{ss5} we plot the
inclination, azimuthal angle and surface density of a disc with a
tidal warp radius of $R_{\rm tid}=5\,R_\star$ and $K=100$.  Once the
inner edge of the disc moves away from the Be star equator, accretion
on to the Be star begins. The disc reaches a steady state that similar
to the one with the zero torque inner boundary condition (shown in
Fig.~\ref{acc5}) but now is closer to alignment with the binary plane
because there is more mass in the disc.

\begin{figure*}
\begin{center}
  \epsfxsize=8.4cm \epsfbox{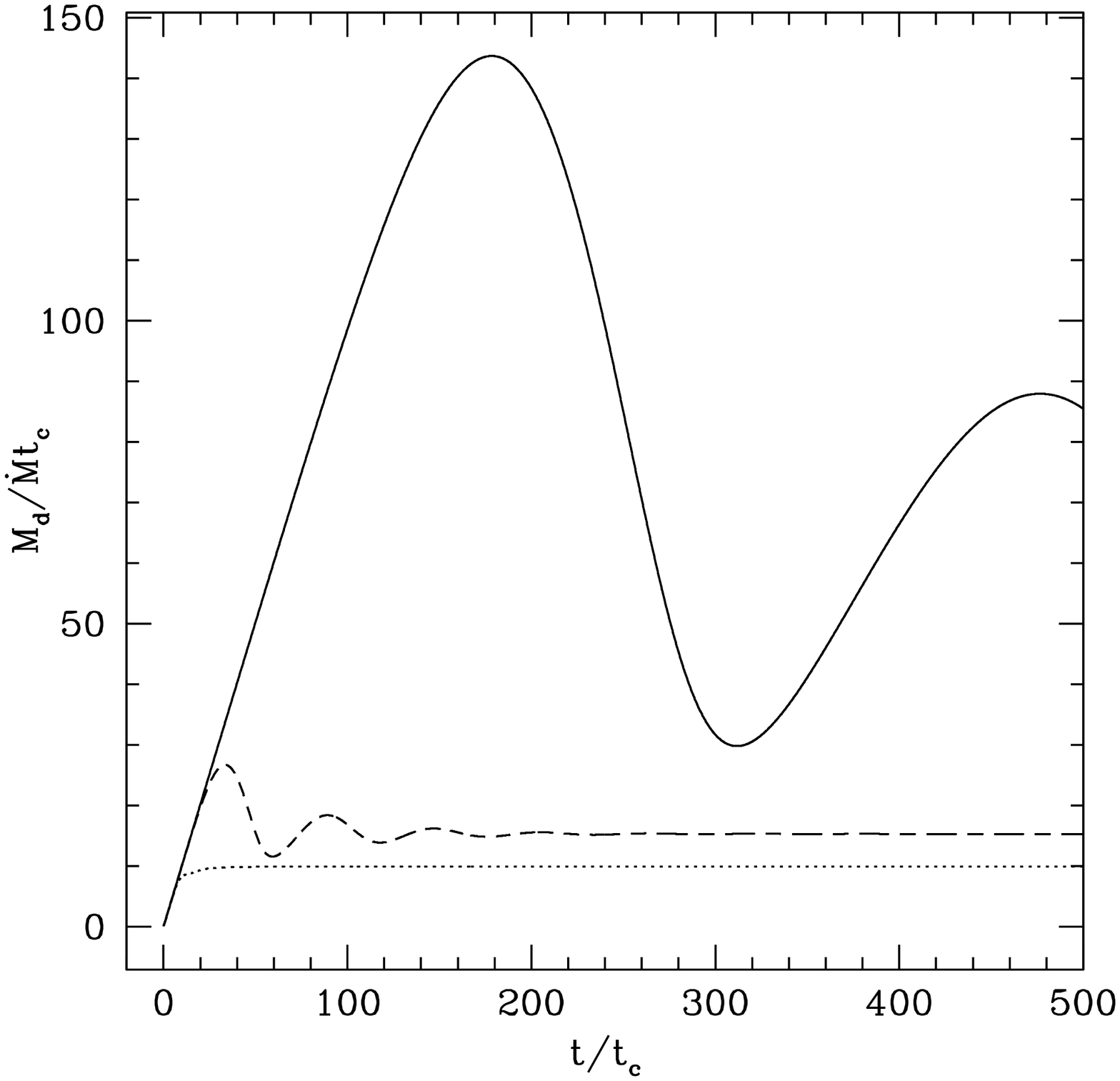}
 \epsfxsize=8.4cm \epsfbox{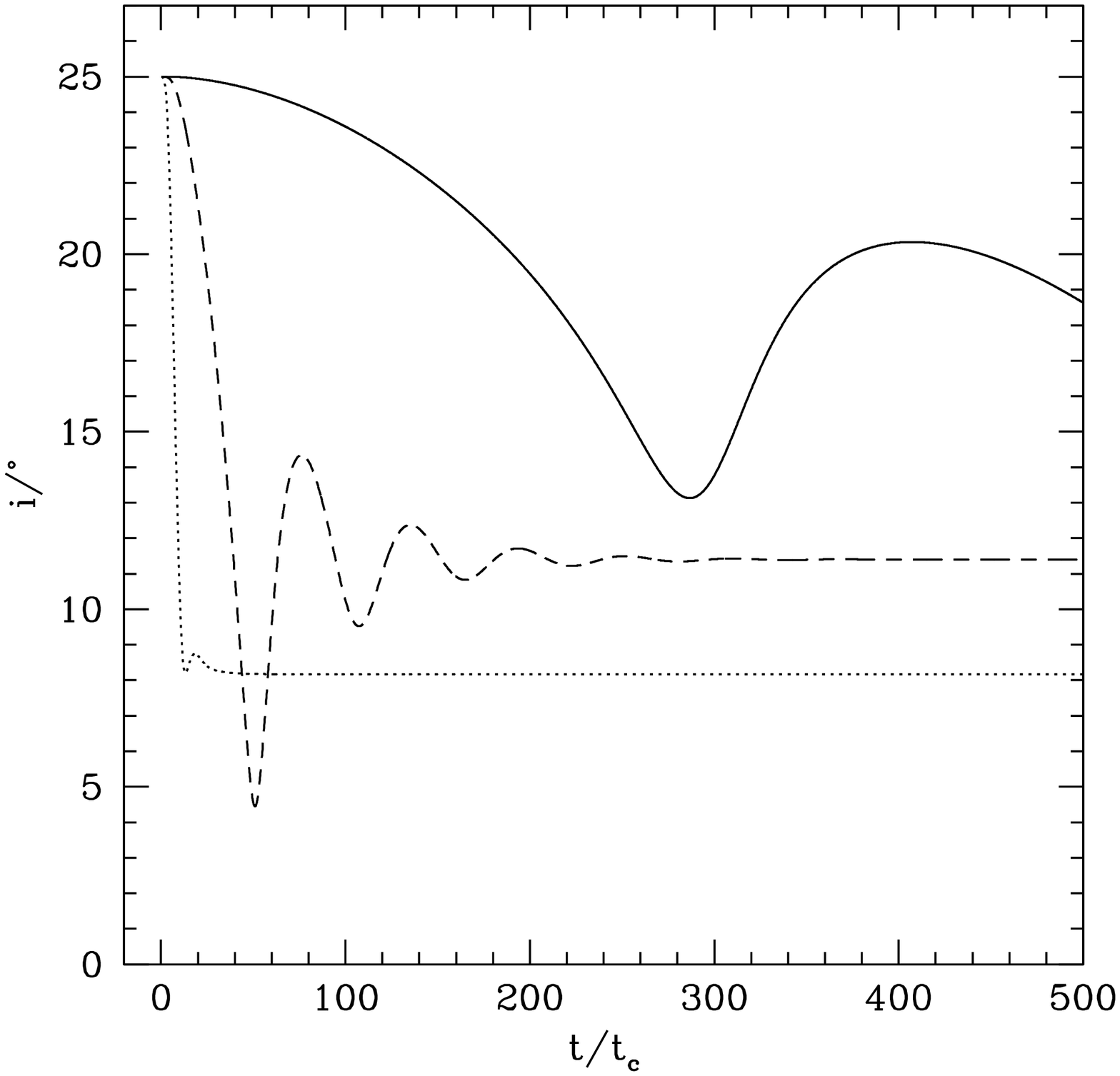}
  \caption[] {Three models with the varying inner boundary condition
    in equation~(\ref{secondbc}) with $K=10$. Left: The evolution of
    the total mass in the disc. Right: The evolution of the
    inclination of the inner edge of the disc. The spin of the Be star
    is at a misalignment of $\theta=25^\circ$ in all three cases. The
    dotted lines show a model with $R_{\rm tid}=5\, R_\star$, the
    dashed lines with $R_{\rm tid}=20\,R_\star$ and the solid lines
    with $R_{\rm tid}=80\,R_\star$.  The time unit is defined in
    equation~(\ref{tc}).}
\label{mass}
\end{center}
\end{figure*}

In Fig.~\ref{mass} we plot the evolution of the mass of the disc
(left) and the inclination of the inner disc edge relative to the
binary orbital plane (right) for discs with $R_{\rm tid}=5$, $20$ and
$80\,R_\star$ and $K=10$.  There is competition between the tidal
torque, the mass addition torque and the magnetic torque which causes
oscillations in the mass and inclination of the inner disc.
Eventually the disc always reaches a warped and twisted steady state.
The larger the tidal warp radius, the larger in both amplitude and
number the oscillations are in mass and inner inclination of the
disc. With a larger warp radius, the steady state mass of the disc is
larger and the inclination of the inner disc is closer to the Be star
equator. The discs shown in Fig.~\ref{mass} have warp radii of $R_{\rm
  tid}=5$, $20$ and $80\,R_\star$ so their corresponding tidal
timescales at the inner edge are $11.1$, $89.4$ and $715\,t_{\rm
  c}$. This is the timescale that the discs reach their steady
state. The oscillations occur on a timescale shorter than this.

Once the disc mass has built up to a steady state, the only way for
the oscillations to reoccur is for the disc mass to be lost or the
accretion on to the disc to slow or cease. For example, if the driving
mechanism behind the mass ejection from the star is pulsation then it
is possible that the oscillations will continue as the accretion on to
the disc varies over time.

\begin{figure*}
\begin{center}
  \epsfxsize=5.5cm \epsfbox{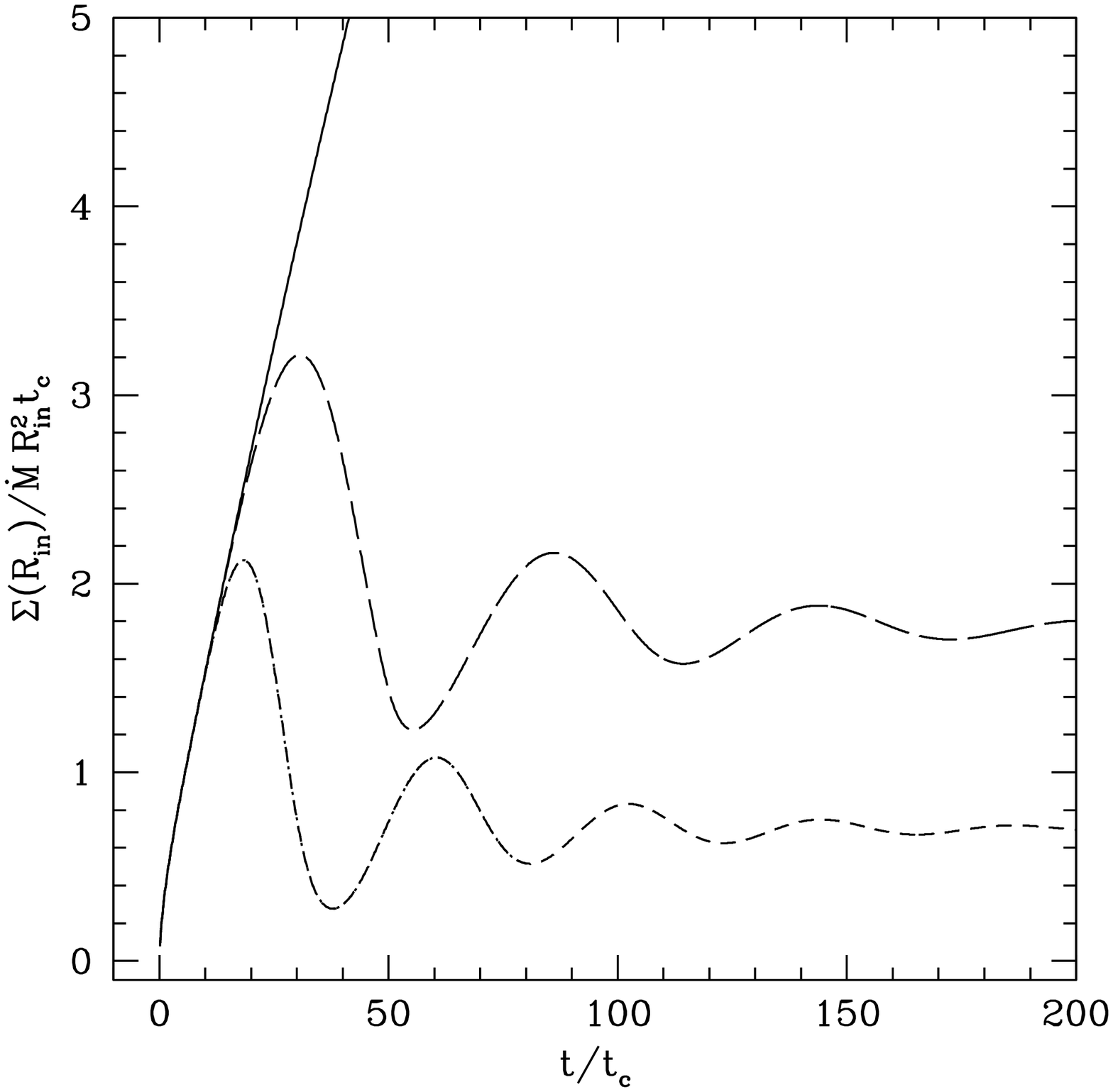}
  \epsfxsize=5.5cm \epsfbox{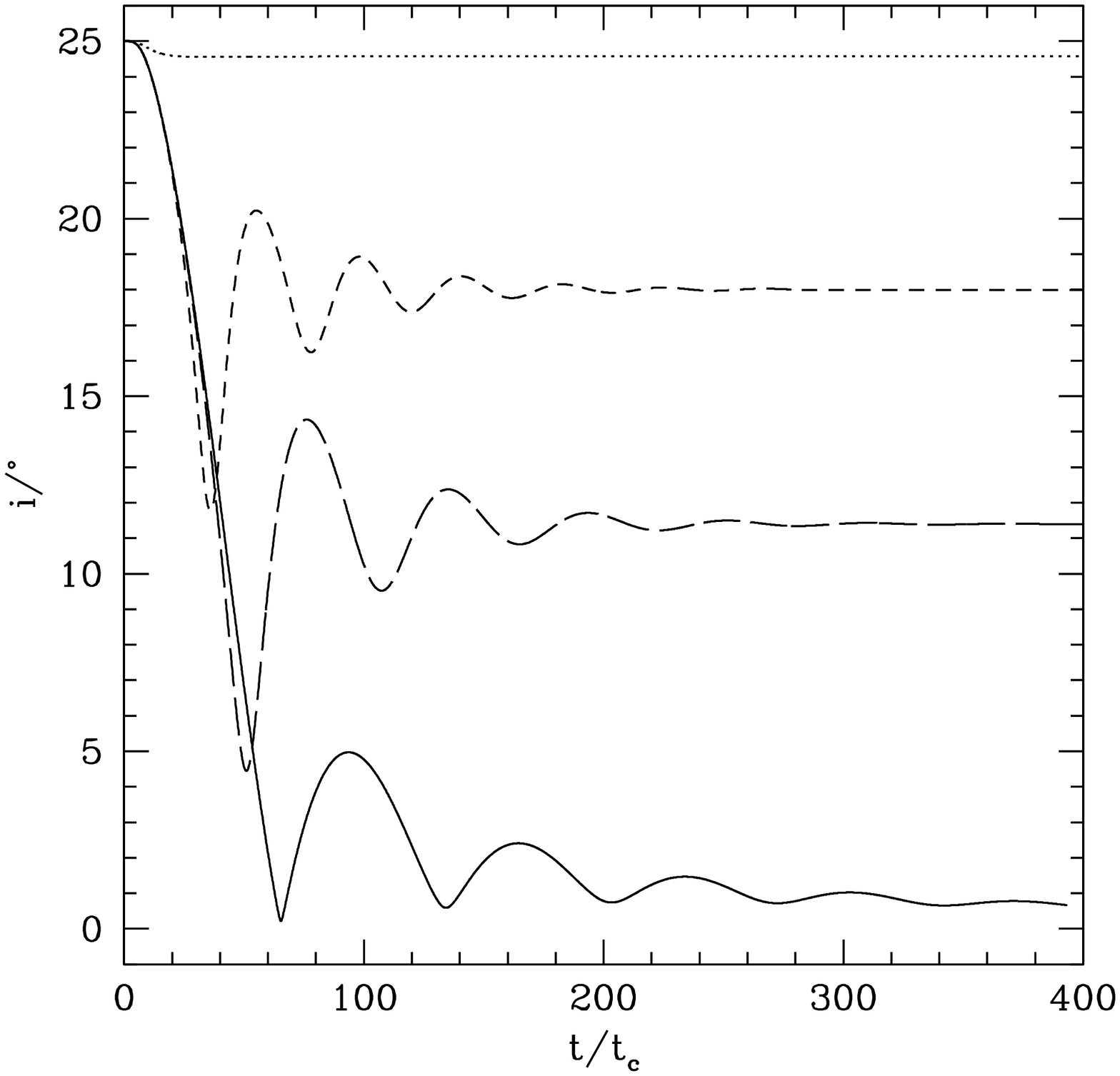}
  \epsfxsize=5.5cm \epsfbox{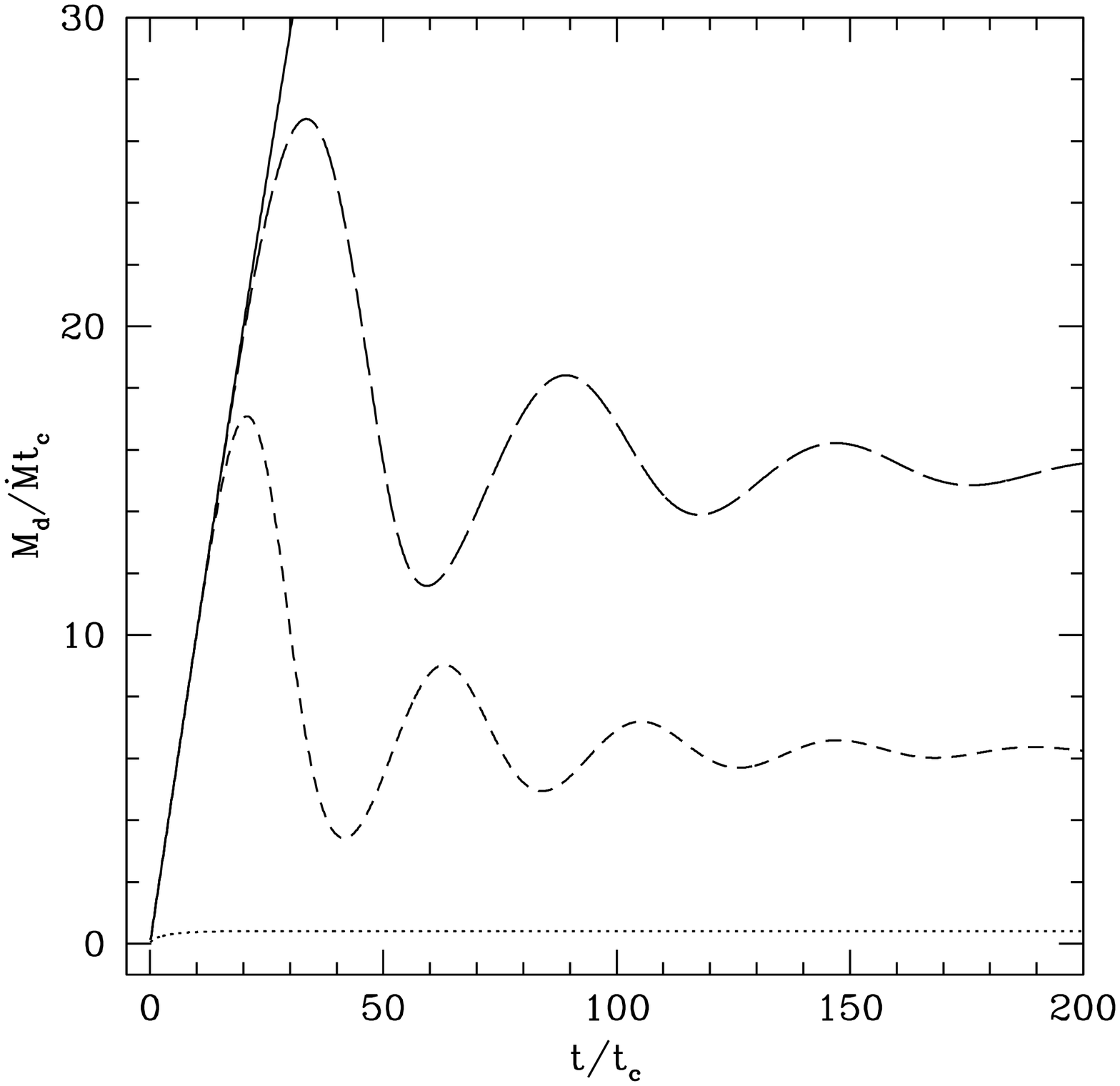}
  \caption[] {Discs with $R_{\rm tid}=20\,R_\star$. Left: The surface
    density at the inner edge of the disc. Middle: The inclination of
    the inner edge of the disc. Right: The total mass of the disc. In
    each plot the dotted line shows a disc with $\Sigma_\star=0$ (not
    shown in the left plot because it is zero). The solid line shows a
    disc with $v_{\rm R}=0$ inner boundary condition. The dashed lines
    show discs with the varying inner boundary condition with $K=10$
    (long-dashed lines) and $K=100$ (short dashed lines).  }
\label{kterm}
\end{center}
\end{figure*}

In Fig.~\ref{kterm} we consider the effect of varying the constant
$K$. We plot the surface density and inclination of the inner parts of
the disc, and the total disc mass with varying $K$ for $R_{\rm
  tid}=20\,R_\star$.  The larger $K$, the smaller the disc mass and
surface density at the inner edge. The larger $K$, the larger the
steady state inclination of the disc to the binary orbit. We also plot
the evolution of the zero radial velocity inner boundary condition and
see that when the tidal warp radius is outside of the disc then there
are small oscillations in inclination of the disc as the disc
approaches the binary orbital plane. Precession of the angular
momentum vector is shown in the bottom right plot of Fig.~\ref{inc2}
also.

\begin{figure}
\begin{center}
  \epsfxsize=8.4cm \epsfbox{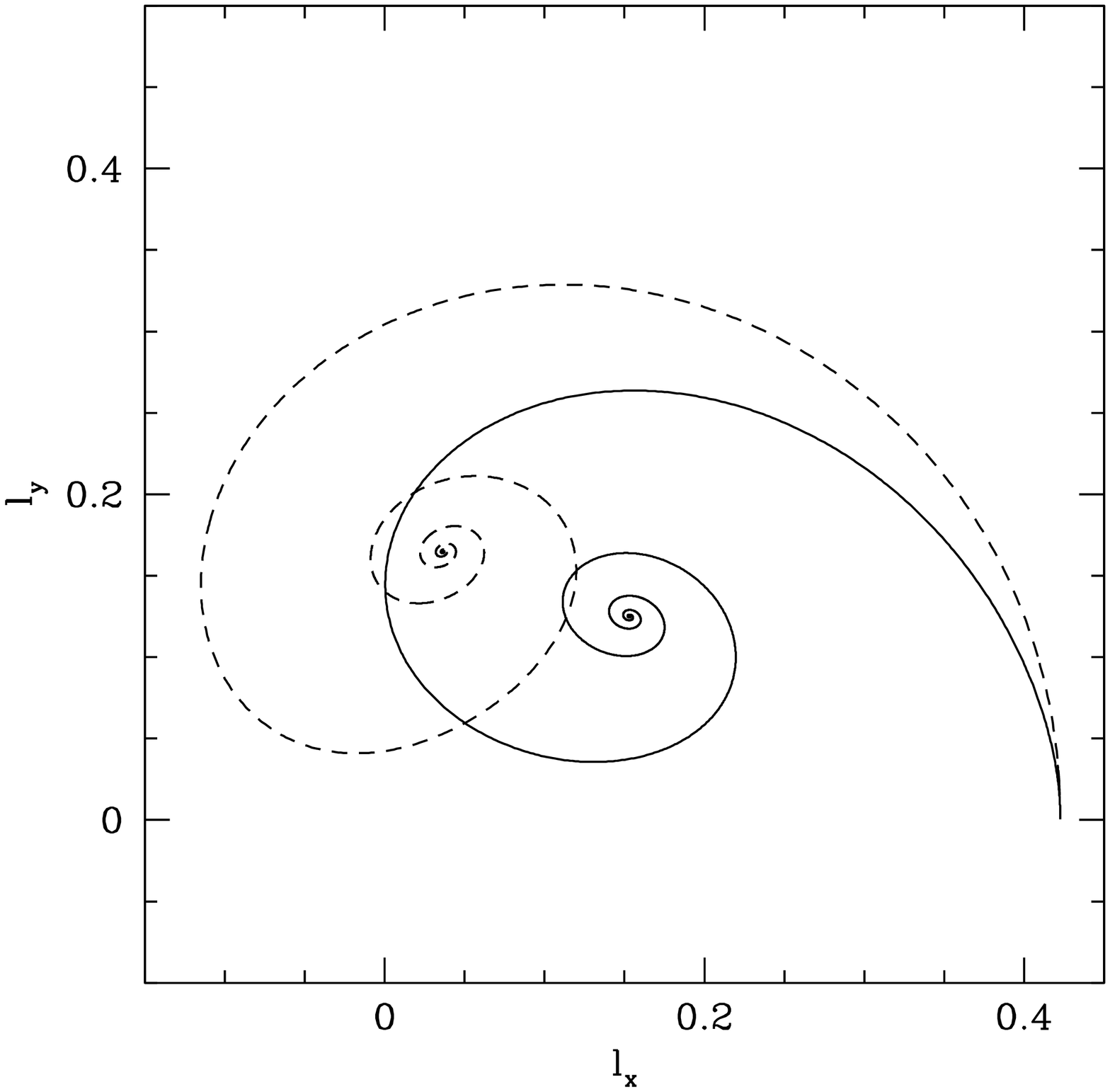}
  \caption[] { The evolution of the inner (solid line) and outer
    (dashed line) angular momenta of the disc with the varying inner
    radial velocity inner boundary condition with $K=10$ and $R_{\rm
      tid}=20\,R_\star$. At $t=0$ the disc begins with $l_y=0$ on the
    right hand side of the plot. }
\label{j}
\end{center}
\end{figure}

In Fig.~\ref{j} we plot the evolution of the angular momentum vector
at the inner and outer edge of the disc for a disc with $R_{\rm
  tid}=20\,R_\star$ and $K=10$ which we can compare with the similar
plot in Fig.~\ref{inc2} which instead had a zero radial velocity inner
boundary condition. We see clearly the disc precesses about the steady
state solution.

\begin{figure}
\begin{center}
  \epsfxsize=8.4cm \epsfbox{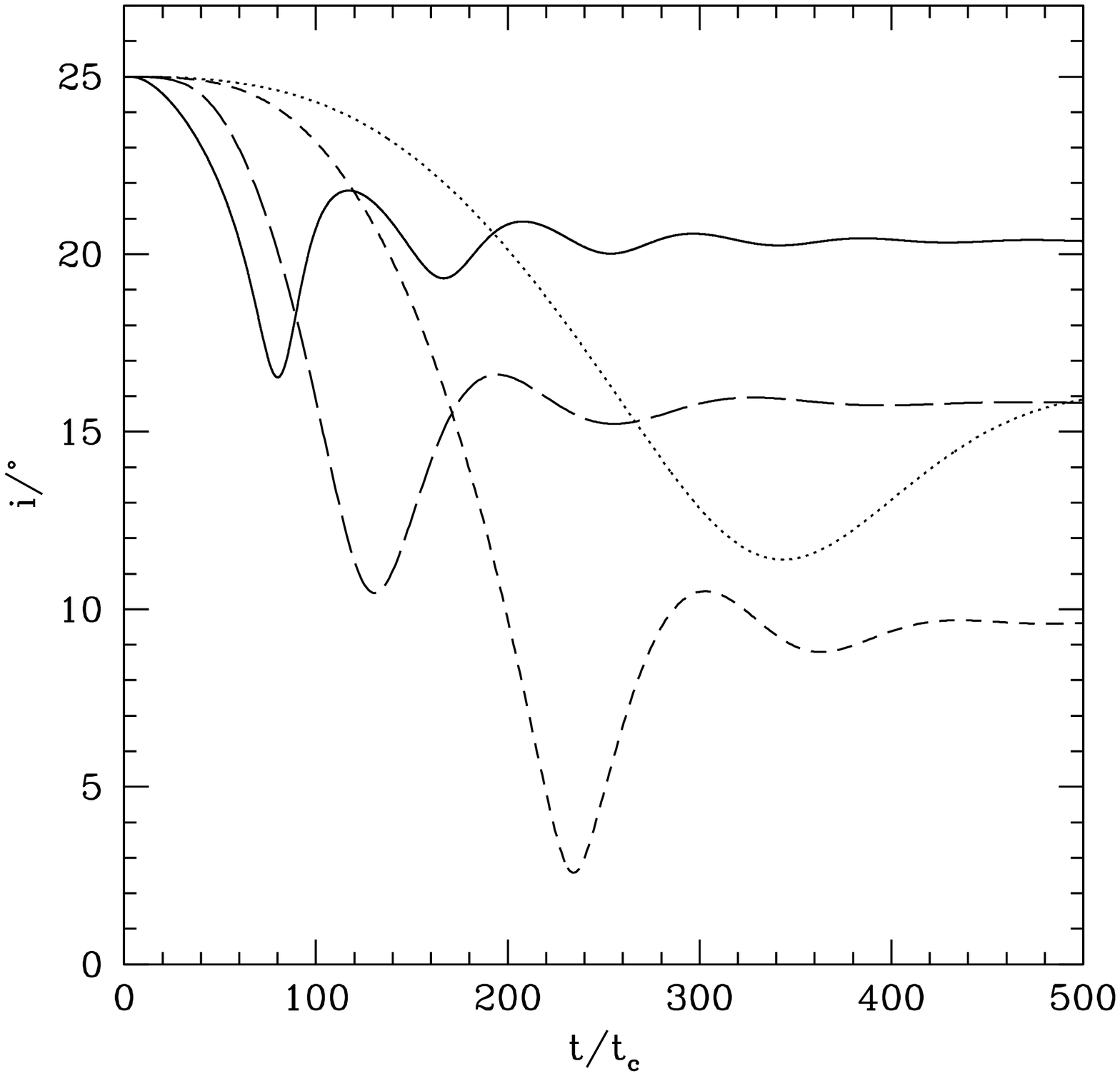}
  \caption[] {The inner disc inclination evolution for models with
    $R_{\rm tid}=40\,R_\star$ and $K=100$. The solid line has
    $\alpha_2/\alpha_1=0.1$ and $n= 4$. The long-dashed line has
    $\alpha_2/\alpha_1=0.1$ and $n=7/2$. The short dashed line has
    $\alpha_2/\alpha_1=0.02$ and $n=7/2$. The dotted line has
    $\alpha_2/\alpha_1=0.1$ and $n= 3$.}
\label{alp}
\end{center}
\end{figure}

In all of the models shown so far, the viscosity radius power law
index is $n=4$ (see equation~\ref{viscosityp}). However,
\cite{waters87} and \cite{dougherty94} matched the infrared excess of
a large sample of Be stars and found the index to be in the range
$2\le n\le 5$. An isothermal disc theoretically requires $n\ge 3.5$
for outflow \citep{porter99}. More recently \cite{jones07} used
hydrodynamical simulations of the disc structure and predicted the
range $3\le n \le 3.5$ in the inner regions of the disc.  In
Fig.~\ref{alp} we show the effect of varying the viscosity parameters
$n$ and $\alpha_2/\alpha_1$. Comparing the solid line (with $n=4$),
the long dashed line (with $n=7/2$) and the dotted line (with $n=3$),
we find the smaller $n$ is the longer the timescale of the
oscillations.  However, qualitatively the results are similar. We note
that when we vary $n$, we change the viscous timescales. In order to
compare models with the same tidal warp radius but varying $n$, the
second viscous timescale at the inner edge of the disc changes as
$t_{\rm c}(n)= t_{\rm c}(n=4)(R_\star/R_{\rm tid})^{4-n}$.  Comparing
the long-dashed line ($\alpha_2/\alpha_1=0.1$) with the short-dashed
line ($\alpha_2/\alpha_1=0.02$), we see that with a smaller value of
$\alpha_2/\alpha_1$ the oscillations become slightly larger and the
steady state is closer to alignment with the binary orbit. However,
the qualitative behaviour of the system is similar with varying
$\alpha_2/\alpha_1$.

We have only considered Be stars that are rotating at break up with
$\Omega_{\rm star}=\Omega_\star$.  We note that if $\Omega_{\rm
  star}<\Omega_\star$ then the oscillation amplitude decreases and the
disc moves to steady state more quickly.

\section{Observed Be Star Binaries}
\label{applications}

\begin{table*}
\begin{center}
\footnotesize
\begin{tabular}{ l l l l l l l l }
  \hline
 Be/B Star Binary & Spectral Type& $P$/d & $e$  & $\theta/^\circ$  \\ \hline
  \hline 
  4U0115+634 (V635 Cas)  &B0.2 Ve  & 24.3 & 0.34 $^{[1]}$  & {\rm unknown} $^{[2]}$ \\
  59 Cyg      & B1e $^{[3]}$      & 28.2  & 0.2 $^{[4]}$  &  {\rm  unknown}  $^{[5]}$ \\
  J0045-7319  &  B1 V  &     51.2      &      0.808 $^{[6]}$     & $25-41\,^{[7],[8],[9]}$   \\
  A0535+262 (V725 Tau)  & B0 III-Ve $^{[10]}$& 111 $^{[11]}$& 0.47 $^{[12]}$ & {\rm unknown} $^{[13]}$ \\
$\zeta$ Tau   &  B2 III    & $133$ $^{[17]}$ &    & {\rm unknown} $^{[18]}$    \\
  0053+604 ($\gamma$ Cas) & B0.5 IVe  & 203.59   & 0.26 $^{[14]}$ &   $\approx 25$ $^{[5]}$  \\
  28 Tau   & B8e  &  218  & 0.6 $^{[15]}$ & 59 $^{[16]}$ \\
  SS 2883  (B1259-63)  & B2e  &   1236.7    & 0.87 $^{[7]}$    & $>$ $55$$\,^{[7],[6]}$    \\
  \hline
 \end{tabular}
\end{center}
\caption[]{Binary Be~stars and binary B stars that have either an
  observed misalignment or inferred warping. We show in column~3 the
  orbital period, $P$, column~4 the eccentricity, $e$ and column~5 the
  misalignment, $\theta$, between the spin of the Be star and the binary
  orbital axis. Where there is no measured misalignment we give
  references for the inferred warped, tilted or precessing disc. 
  $^{[1]}$\protect\cite{rappaport78} 
$^{[2]}$\protect\cite{neg01}
 $^{[3]}$\protect\cite{harmanec02} 
$^{[4]}$\protect\cite{rivinius00}
$^{[5]}$\protect\cite{Hummel98}
$^{[6]}$\protect\cite{wex98}
 $^{[7]}$\protect\cite{hughes99} 
  $^{[8]}$\protect\cite{kaspi96}
 $^{[9]}$\protect\cite{L95}
 $^{[10]}$\protect\cite{steele98} 
$^{[11]}$\protect\cite{hutchings84}
$^{[12]}$\protect\cite{neg00}
  $^{[13]}$\protect\cite{Larionov01} 
  $^{[14]}$\protect\cite{harmanec00}  
  $^{[15]}$\protect\cite{katahira96}
$^{[16]}$\protect\cite{hirata07}
$^{[17]}$\protect\cite{harmanec84}
$^{[18]}$\protect\cite{schaefer10}}
\label{tab}
\end{table*}

In Table~\ref{tab} we show some Be and B star binaries that have been
suggested to have a warped or tilted disc or have a measured
misalignment between the Be star spin and the binary orbital
axis. There is one B star binary because it has a measured
misalignment and it is thought that B stars become Be stars and vice
versa.  We show the spectral type, the period, $P$, the eccentricity,
$e$, and the measured inclination, $\theta$, between the spin of the
Be star and the axis of rotation of the disc.  Even if there is no
observed misalignment for a Be star system, the formation mechanism of
the companion neutron star by an asymmetric supernova did most
probably produce a misalignment \citep{MTP09}. There is a large range
of orbital periods and eccentricities in the Be/B star binaries.

Ideally we should make an individual model of each system in order to
draw firm conclusions about its evolution. However, in this Section we
just consider the general properties of the Be star systems. We choose
the mass of the Be star according to its spectral type
\citep[see][]{eggleton89} and then find the corresponding stellar
radius from the mass \citep[e.g.][]{tout96}. We list the properties in
Table~\ref{tab2}.

In Table~\ref{tab3} we show the disc properties in each Be star
binary.  We find the semi-major axis of the orbit from
equation~(\ref{semimajor}) and then the periastron separation is
$p=(1-e)a$. We find the average separation with
equation~(\ref{r3}). The tidal warp radius is found with
equation~(\ref{rtid2}) with $n=4$ and then the tidal timescale with
equation~(\ref{tidtime}). The tidal timescale tells us the timescale
on which the disc reaches steady state, as shown in
Section~\ref{accretion}. Any oscillations of the disc mass and inner
inclination of the disc occur on a timescale shorter than this.

\begin{table}
\begin{center}
\footnotesize
\begin{tabular}{ l l l l l l l l }
  \hline
 Be Star System Properties\\ \hline
  \hline 
Be Star \\
{        }Spectral Type & B0 & B1 & B2 & B8 \\
{        }Mass $M_1/\rm M_\odot$  & 17 & 15  & 12  & $2.9$\\
{        }Radius $R_\star/R_\odot$ & 10 & 8.5 & 7& 3.7 \\
\hline
Mass of Neutron Star, $M_2$ & $1.4 \,\rm M_\odot$ \\
Disc Scale Height $(H_\star/R_\star)$ & 0.04 \\
  \hline
 \end{tabular}
\end{center}
\caption[]{Properties of the Be star systems we choose. We use the B0
  star parameters throughout the work and the other parameters for
  Table~\ref{tab3}.}
\label{tab2}
\end{table}

\begin{table*}
\begin{center}
\footnotesize
\begin{tabular}{ l l l l l l l l }
  \hline
  Be/B Star Binary& $a/R_\star$ & $p/R_\star$  & $\bar R_{\rm b}/R_\star$ 
&  $R_{\rm tid}/R_\star$    & $t_{\rm tid*}/{\,\rm yr}$  \\ \hline
  \hline 
  4U0115+634 (V635 Cas)  & 9.3 & 6.2 & 8.8    &  8.2 &  2.1        \\
  59 Cyg       & 11.7   & 9.3   & 11.4   &  12.7  &  3.4            \\
  J0045-7319  & 17.4 & 3.3 & 10.2     & 10.2 & 3.3                \\
  A0535+262 (V725 Tau)  & 25.7 & 13.6 & 22.6    &  54.4 & 36.5     \\
  0053+604 ($\gamma$ Cas)  & 38.5  & 28.5  &  37.2 & 146.6 & 161.4  \\
  28 Tau  &   67.1  & 26.8  & 53.6  & 93.7 & 44.9                  \\
  SS 2883  (B1259-63)  & 165 & 21.4  &   81.2    & 554.5 &  825.2  \\
  \hline
 \end{tabular}
\end{center}
\caption[]{The binary Be/B star binaries shown in Table~\ref{tab} that
  have measured period and eccentricity. Here we compute the binary
  semi-major axis, $a$, the periastron separation, $p$, the average
  radius, $\bar R_{\rm b}$, the tidal warp radius, $R_{\rm tid}$ and
  the tidal timescale at $R_\star$. We choose the mass and radius of
  the Be star system depending on its spectral type from
  Table~\ref{tab}.}
\label{tab3}
\end{table*}


The systems V635 Cas and 59 Cyg should have an evolution similar to
that described by Fig.~\ref{ss5} as they both have small tidal warp
radii. The mass and inner disc inclination evolution is similar to the
dotted line in Fig.~\ref{mass}.  They have tidal timescales of a few
years and so should reach a steady state disc relatively quickly. The
oscillations in mass and inner disc inclination should be minimal if
the accretion rate on to the Be star decretion disc is
constant. However, if the mass ejection rate from the Be star is not
constant then the warping and precession could continue over time.

\cite{neg01} observed the Be/X-ray transient 4U0115+63/V635 Cas in
optical, infrared and X-rays and that the disc warps, tilts and starts
precessing. They found a quasi-period cycle with a period of $3-5\,\rm
yr$ on which the Be star disc forms, grows and is lost.  They observed
two Type~II outbursts before the disc became faint. The H$\alpha$ line
profile in this system is usually double peaked. However, just before
the outbursts, the line became single peaked and this was sometimes
followed by a shell event. They assume that the warping is because the
disc becomes unstable to radiation driven warping. However, in such a
short period binary with only a moderate eccentricity the tidal
torques will certainly cause the disc to warp and precess if the
system is misaligned.  As the warped disc precesses, the line profile
changes shape.  \cite{neg01} suggest that the warped parts of the disc
will have a strong interaction with a fast stellar wind and stellar
irradiation and so a large amount of gas falls on to the neutron star
causing a Type~II outburst and then the disc becomes faint because it
has lost a lot of mass. We have found that the tidal timescale in the
inner parts of the disc is about 2.1 years. This is the timescale on
which the disc would reach steady state or else its maximum warping if
the accretion on to the disc from the star is at a constant
rate. However, with a varying mass ejection rate from the Be star, the
disc will continue to warp and precess as the mass in the disc varies.


The B-star binary PSR J0045--7319 has a spin-orbit misalignment
suggested by its orbital plane precession \citep{kaspi96,L95}. In this
case the B star rotates retrogradely with respect to the orbit
\citep{L96a}. The system has a tidal timescale of $3.3\,\rm yr$. This
system should show oscillations in its mass and inner disc inclination
on the timescale of a year or so if a disc was ejected.


\cite{haigh04} found that the disc in A0535+26 grows and decays on a
quasi-cycle period of about $1500\,\rm d$ by observing the change in
quantised, IR excess flux states.  The system has also shown type II
outbursts in 1994 and 2005 \citep{coe06}.  Because the system has only
a moderate eccentricity, unless it has a very large misalignment
angle, the disc will be truncated by tidal torques before it can reach
the orbit of the neutron star.  We find the tidal timescale in this
system to be around 37 years and because the tidal warp radius is
large, oscillations in mass and inclination should occur on a
timescale of around the order of 10 years.


The Be star $\zeta$ Tau has a companion that is yet to be
detected. Because the eccentricity is unknown we do not compute the
tidal timescales but note that the system may be similar to
A0535+26. Recently \cite{schaefer10} used interferometry to observe
the disc and suggested that the tilt of the disc is precessing. 


$\gamma$ Cas showed two successive shell phases over a period of 6
years \citep{Hummel98}. We find the tidal timescale at the inner edge
of the disc to be $161$ years. The tidal warp radius is very large, at
$147\,R_\star$. This implies that the system will undergo large
oscillations in disc mass and inner disc inclination as the disc mass
builds up.


\cite{hirata07} use polarimetry to observe the precession in 28 Tau
(Pleione). Over the course of a hundred years, the star has changed
from Be $\rightarrow$ B, shell $\rightarrow$ Be, shell $\rightarrow$
Be with an activity cycle of $35\,\rm yr$. It is currently in the Be
phase. They find that over the time period 1974 to 2003 the disc axis
precessed from $60^\circ$ to $130^\circ$. With the H$\alpha$ profile
they suggest that the disc inclination has also changed
drastically. They found the precession period of 28 Tau to be
$80.5\,\rm yr$ but assumed a constant precession rate. Our model
predicts that the precession period is $45\,\rm yr$.  The rigid body
precession model of \cite{larwood98} found that the precession period
is $270\,\rm yr$ and $70\,\rm yr$ for disc radii of $10\,R_\star$ and
$25\,R_\star$ respectively, also in reasonable agreement.


The tidal torque from the companion will have an effect on all of the
Be stars in listed in Table~\ref{tab3} apart from maybe SS2883 which
has the longest orbital period and a very large eccentricity. The
timescale on which the averaged tidal torque acts is very long.
Because of the large eccentricity, a one armed spiral structure may
form at periastron passage \citep{hayasaki04,hayasaki05}.  The fact
that tidal torques are negligible in this system has been noted before
by \cite{kochanek93,manchester95,L95}. \cite{wex98} suggests that the
solution is a precessing orbit caused by the quadrupole moment of the
tilted companion star.

\section{Conclusions}

We find that in Be star discs the torques from a neutron star
companion are important and the discs can become warped and twisted if
the system is misaligned (except in the systems with the longest
periods). Even though Be star's have eccentric orbits with their
companions, the orbital period is short compared the timescale on
which the tidal torque acts and so we can average the tidal torque
over an orbital period. The discs are truncated at a radius smaller
than the periastron separation of the binary providing the the
eccentricity is not close to 1 and the misalignment is not too
large. The size of the disc decreases with increasing eccentricity of
the binary orbit. In most systems, it is unlikely that the neutron
star will pass through the disc and so Type II outbursts are more
likely linked to the warping of the disc.

We have made numerical models of Be star discs that are assumed to be
truncated by the companion and considered solutions with three
different inner boundary conditions that approximate the magnetic
torque from the Be star.  First we considered a magnetic torque strong
enough to prevent all accretion back on to the Be star, or zero radial
velocity.  In this case the disc mass increases linearly in time and
the disc tends towards complete alignment with the binary orbital
plane. Secondly we considered a disc with a zero torque inner boundary
condition normally associated with accretion discs. The surface
density at the inner edge of the disc is zero and the disc has a small
mass. The disc moves towards a steady state solution with the inner
parts close to alignment with the equator of the Be star and the outer
parts become warped towards the binary plane but shows little
precession.

Finally we considered a magnetic torque that is strongest at the
equator of the Be star (where it rotates fastest) and decreases as the
disc moves away from there. The corresponding boundary condition
varies the radial velocity at the inner edge of the disc with the
inclination to the Be star equator.  The mass addition and tidal
torques oscillate in dominance as the surface density varies.  A
warped and twisted steady state is reached once the oscillations die
away on the tidal timescale at the inner edge of the disc which is of
the order of a year up to a few hundred years (see
Table~\ref{tab3}). On a timescale shorter than this the disc mass and
inclination oscillate if the tidal warp radius is large. We suggest
that these oscillations could be the cause of the shell events. The
inclination and mass of the disc change on a timescale of a few, to a
few tens of years until the disc reaches its steady state. If Type~II
X-ray outbursts are caused by a warped disc interacting with a fast
stellar wind then the tidally warped disc solution could provide
warping on a timescale similar to those observed. If the disc
undergoes periodic mass ejections followed by a decay of the disc
mass, then the angular momentum distribution can be reset to being
small and aligned with the Be star spin. As the disc mass builds up
each time the oscillations will be observed.

Observers should try to measure the misalignment between the spin of
the Be star and the binary orbital plane of more Be star systems. We
note that the inclination of the disc is not necessarily aligned with
either the binary orbit or the spin of the Be star and so the the two
planes need to be measured independently of the disc inclination.

\section*{Acknowledgements}
 We thank Phil Charles for useful conversations. RGM thanks the Space
 Telescope Science Institute for a Giacconi Fellowship

\appendix

\section{Inner boundary conditions}

Here we consider the torque on the inner parts of the disc from the
rotating Be star. The torque may be created by the magnetic field of
the star \citep[e.g.,][]{livio92}. We need only consider a flat disc
to investigate the inner surface density boundary condition because we
have chosen the angular momentum to satisfy $\partial \bm{l}/\partial
R=0$.  The surface density evolution equation for a flat disc with a
source of angular momentum is
\begin{align}
\frac{\partial \Sigma}{\partial t}= & \,\,
\frac{1}{R}\frac{\partial}{\partial R}\left[ 3
  R^\frac{1}{2}\frac{\partial}{\partial R}\left(\nu_1 \Sigma
  R^\frac{1}{2}\right)\right]\cr &  -\frac{1}{R}\frac{\partial}{\partial
  R}\left[\frac{2\Lambda \Sigma R^\frac{3}{2}}{\sqrt{GM_1}}\right],
\label{source}
\end{align}
\citep{pringle91}, where the input rate of angular momentum per unit
mass is
\begin{equation}
\Lambda=\frac{GM_1}{R}f
\end{equation}
and $f(R)\le 1$ measures the strength of the torque
\citep{pringle91}. The function $f$ is a strongly decreasing function
of radius \citep{papaloizou84,pringle91}.  The radial velocity in
the disc is
\begin{equation}
v_{\rm R}=\frac{2 \sqrt{GM_1}}{R^\frac{1}{2}}f-\frac{3}{\Sigma R^\frac{1}{2}}\frac{\partial}{\partial R}\left(\nu_1\Sigma R^\frac{1}{2}\right).
\label{vf}
\end{equation}
We can rearrange this to find 
\begin{equation}
f=\frac{R^\frac{1}{2}}{2(GM_1)^\frac{1}{2}}\left[
v_{\rm R}+\frac{3}{\Sigma R^\frac{1}{2}}\frac{\partial }{\partial R}\left(\nu_1\Sigma R^\frac{1}{2}\right)\right].
\label{f}
\end{equation}

\subsection{Zero Radial Velocity}
\label{app}

First we consider a torque that is strong enough to prevent all
accretion back on to the star. This is equivalent to the inner
boundary condition $v_{\rm R}=0$ at $R=R_\star$ no matter what the
inclination of the inner parts of the disc relative to the Be star
spin.  In this case with equation~(\ref{vf}) we find the torque must
be such that
\begin{equation}
f=\frac{3}{2\sqrt{GM_1}\Sigma}\frac{\partial}{\partial R}\left(\nu_1 \Sigma R^\frac{1}{2}\right)
\label{f0}
\end{equation}
and because $f$ falls off quickly with radius, this is approximately
equivalent to $f=0$ and
\begin{equation}
\frac{\partial}{\partial R}\left(\nu_1 \Sigma R^\frac{1}{2}\right) \Bigl\lvert_{R=R_\star}=0.
\end{equation}
We can replace the source term in equation~(\ref{source}) with this
boundary condition.

\subsection{Zero Torque}

Next we consider a disc with a zero torque inner boundary
condition. We can choose $f=0$ and then the torque on the disc is the
usual viscous torque
\begin{equation}
T=-3 \pi \nu_1\Sigma (GM_1R)^\frac{1}{2}.
\end{equation}
For zero torque at the inner edge we have $\Sigma=0$ at $R=R_\star$.

\subsection{Varying Inner Radial Velocity}

The Be star rotates fastest at the equator. The third torque we
consider varies with the inclination of the inner parts of the
disc. The torque is strongest at the equator and falls off as the disc
moves away from the equator.  We choose a torque of the form
\begin{equation}
f=C\left( \frac{R}{R_\star}\right)^{-10} \exp\left[\frac{-(i_\star - \theta)^2}{\delta^2}\right],
\label{f2}
\end{equation}
where $C$ and $\delta$ are constant. We can vary the rate at which the
torque falls off away from the equator of the Be star and the torque
falls off quickly with radius from the star.

If the constant, $C$, is large enough, this is equivalent to an inner
boundary condition that takes $v_{\rm R}=0$ only at the equator, when
the star rotates at break up, but allows for accretion on to the Be
star as the inclination of the inner disc moves away from the equator
of the Be star. At the equator $i_\star=\theta$. As the inner disc
inclination moves towards the pole, the viscous torque tends to zero
and $\Sigma_\star=0$ and $v_{\rm R}\rightarrow \infty$. Hence we can
parametrise the radial velocity at the inner edge of the disc as
\begin{equation}
v_{\rm R\star}=-3K \frac{\nu_{1\star}}{R_\star}\left[\frac{\Omega_\star/\Omega_{\rm star}}{\cos (i_\star-\theta)}-1\right],
\end{equation}
where $K$ is a dimensionless constant that we can vary. We cannot
determine $K$ without looking in more detail at the boundary layer
between the disc and the star. The rotation rate of the Be star is
$\Omega_{\rm star}$ and the Keplerian rotation rate of the inner parts
of the disc is $\Omega_\star=\Omega(R_\star)$. 
The boundary condition at the inner edge of the disc is
\begin{equation}
\frac{\partial}{\partial R}(\nu_{1} \Sigma R^\frac{1}{2})\Bigl\lvert_{R=R_\star}=
K\frac{\nu_{1\star}\Sigma_\star }{R_\star^{\frac{1}{2}}}\left[\frac{\Omega_\star/\Omega_{\rm star}}{\cos(i_\star-\theta)}-1\right]
\end{equation}
for $\cos(i_\star-\theta )\ne0$.  If the star is rotating at break up
then $\Omega_{\rm star}=\Omega_\star$.

\begin{figure}
\begin{center}
  \epsfxsize=8.4cm \epsfbox{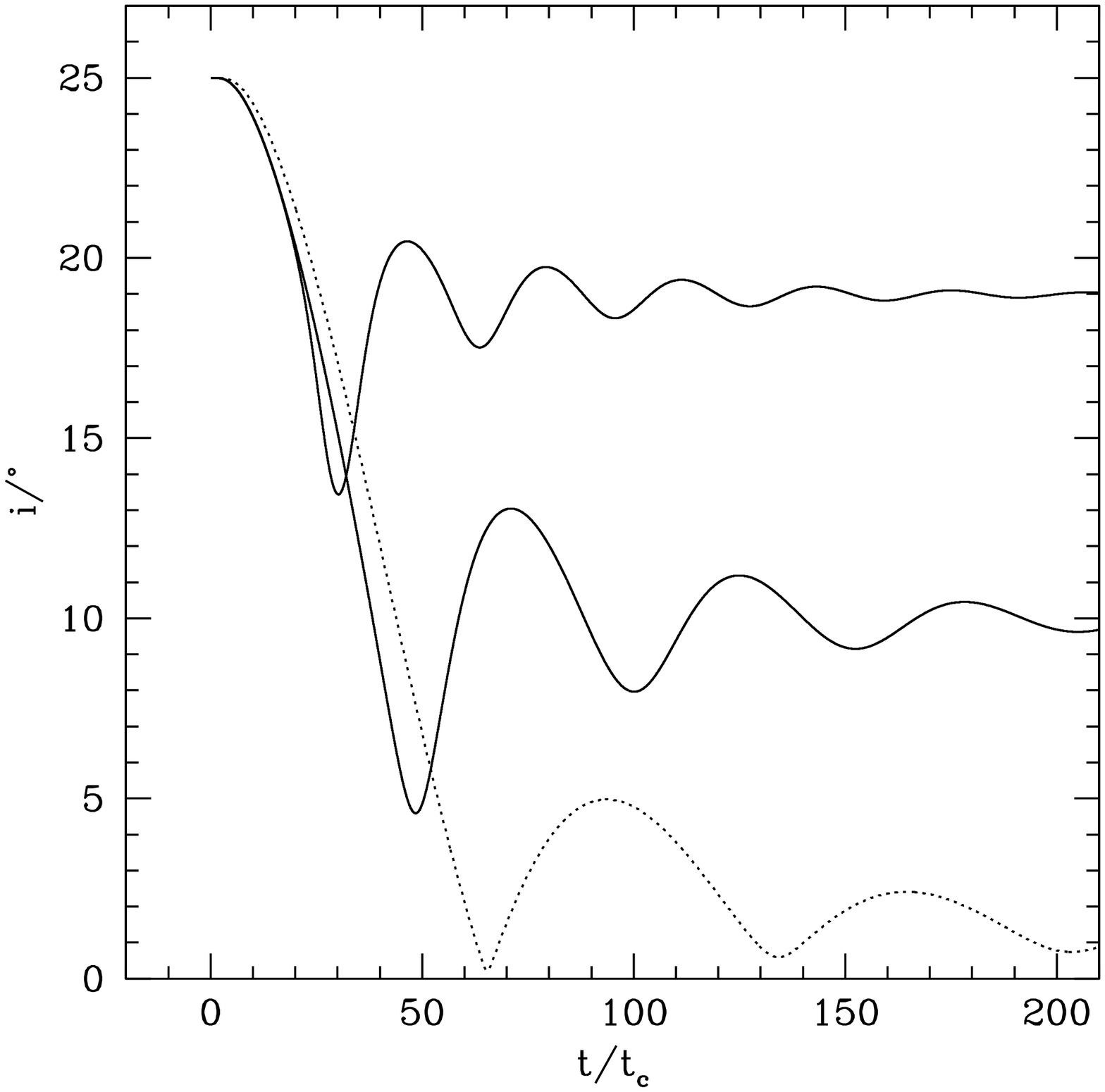}
  \caption[] { A disc with $R_{\rm tid}=20\,R_\star$ including the
    additional torque term given in equation~\ref{f2} and an inner
    boundary condition $\Sigma(R_\star)=0$. We take the constant
    $C=15$ and the upper solid line has $\delta=0.1$ and the lower
    line has $\delta=0.4$. The dotted line shows the case with the the
    inner boundary condition $v_{\rm R}=0$ for comparison (the solid
    line in the middle plot of Fig.~\ref{kterm}). }
\label{torquewidth}
\end{center}
\end{figure}

In Fig.~\ref{torquewidth} we plot the inclination of the inner edge of
the disc as a function of time for a disc with $R_{\rm tid}=20\,
R_\star$ including the varying torque in equation~(\ref{f2}) and an
inner boundary condition of $\Sigma(R_\star)=0$. We see that this
torque does indeed reproduce a similar behaviour to that with the
varying inner boundary condition that we describe in
Section~\ref{accretion}. The inner boundary condition is an
approximation to the torque from a Be star that varies with height
above the equator. However, the constants, $C$, $\delta$ and $K$ can
only be determined by looking in more detail at the magnetic torque
term.

\label{lastpage}
\end{document}